\documentclass[longauth]{aa} 
\usepackage{graphicx}
\usepackage{txfonts}
\def\arcsec{$^{\prime\prime}$}
\def\arcmin{$^{\prime}$}
\def\deg{$^{\circ}$}

\newcommand {\kss} {km~s$^{-1}$}
\newcommand {\msun} {$h^{-1} \  M_{\odot} \;$}

\newcommand{\chandra}{{\sl Chandra}}
\newcommand{\rosat}{{\sl ROSAT}}
\begin{document}
   \title{Sardinia Radio Telescope observations of Abell~194}

   \subtitle{The intra-cluster magnetic field power spectrum}

   \author{F. Govoni\inst{1}
          \and
          M. Murgia\inst{1}
          \and  
          V. Vacca\inst{1}
          \and
          F. Loi\inst{1,2}   
          \and
          M. Girardi\inst{3,4}
          \and
          F. Gastaldello\inst{5,6}
          \and 
          G. Giovannini\inst{7,8}
          \and 
          L. Feretti\inst{7}
          \and 
          R. Paladino\inst{7}
          \and 
          E. Carretti\inst{1} 
          \and
          R. Concu\inst{1}
          \and
          A. Melis\inst{1}
          \and
          S. Poppi\inst{1}
          \and 
          G. Valente\inst{9,1} 
          \and
          G. Bernardi\inst{10,11}
         \and
          A. Bonafede\inst{7,12}
          \and
          W. Boschin\inst{13,14,15}
         \and
          M. Brienza\inst{16,17}
         \and
          T.E. Clarke\inst{18}
         \and
          S. Colafrancesco\inst{19}
         \and
          F. de Gasperin\inst{20}
         \and
          D. Eckert\inst{21}
         \and
          T. A. En{\ss}lin\inst{22}
         \and
          C. Ferrari\inst{23}
         \and
          L. Gregorini\inst{7} 
         \and
          M. Johnston-Hollitt\inst{24}
         \and
          H. Junklewitz\inst{25}
         \and
          E. Orr{\`u}\inst{16}
          \and 
          P. Parma\inst{7}
         \and
          R. Perley\inst{26}
         \and
          M. Rossetti\inst{5}
         \and
          G.B Taylor\inst{27}     
         \and
          F. Vazza\inst{7,12} 
          }

   \offprints{F. Govoni, email fgovoni@oa-cagliari.inaf.it}

   \institute{INAF - Osservatorio Astronomico di Cagliari
              Via della Scienza 5, I--09047 Selargius (CA), Italy
           \and 
              Dip. di Fisica, Universit\`a degli Studi di Cagliari, Strada Prov.le Monserrato-Sestu Km 0.700, I-09042 Monserrato (CA), Italy
           \and  
              Dip. di Fisica, Universit\`a degli Studi di Trieste - Sezione di Astronomia, via Tiepolo 11, I-34143
              Trieste, Italy 
           \and
              INAF - Osservatorio Astronomico di Trieste,
              via Tiepolo 11, I-34143 Trieste, Italy
           \and 
              INAF - IASF Milano, Via Bassini 15, I-20133 Milano, Italy 
           \and 
              Dep. of Physics and Astronomy, University of California at Irvine, 
              4129 Frederick Reines Hall, Irvine, CA 92697-4575, USA
           \and
              INAF - Istituto di Radioastronomia, Bologna 
              Via Gobetti 101, I--40129 Bologna, Italy
           \and 
              Dip. di Fisica e Astronomia, Universit\`a degli Studi di Bologna, Viale Berti Pichat 6/2, I--40127 Bologna, Italy
           \and
              Agenzia Spaziale Italiana (ASI), Roma
           \and 
              SKA SA, 3rd Floor, The Park, Park Road, Pinelands 7405, The Cape Town, South Africa 
           \and
              Department of Physics and Electronics, Rhodes University, PO Box 94, Grahamstown 6140, South Africa
           \and
              Hamburger Sternwarte, Universit\"at Hamburg, Gojenbergsweg 112, 21029, Hamburg, Germany
           \and 
              Fundaci\'on G. Galilei - INAF TNG, 
              Rambla J. A. Fern\'andez P\'erez 7, E-38712 Bre\~na Baja (La Palma), Spain 
           \and 
              Instituto de Astrof\'{\i}sica de Canarias, 
              C/V\'{\i}a L\'actea s/n, E-38205 La Laguna (Tenerife), Spain
           \and 
              Dep. de Astrof\'{\i}sica, Univ. de La Laguna, 
              Av. del Astrof\'{\i}sico Francisco S\'anchez s/n, E-38205 La Laguna (Tenerife), Spain
          \and 
              ASTRON, the Netherlands Institute for Radio Astronomy, Postbus 2, 7990 AA, Dwingeloo, The Netherlands
          \and
             Kapteyn Astronomical Institute, Rijksuniversiteit Groningen, Landleven 12, 9747 AD Groningen, The Netherlands
          \and
              Naval Research Laboratory, Washington, District of Columbia 20375, USA 
          \and
              School of Physics, University of the Witwatersrand, Private Bag 3, 2050, Johannesburg, South Africa 
          \and 
              University of Leiden, Rapenburg 70, 2311 EZ Leiden, the Netherlands
          \and 
              Astronomy Department, University of Geneva, 16 ch. d'Ecogia, CH-1290 Versoix, Switzerland 
          \and
              Max Planck Institut f\"{u}r Astrophysik, Karl-Schwarzschild-Str.1, D-85740 Garching, Germany
          \and 
              Laboratoire Lagrange, UCA, OCA, CNRS, Blvd de l'Observatoire, CS 34229, 06304 Nice cedex 4, France
          \and 
              School of Chemical \& Physical Sciences, 
              Victoria University of Wellington, PO Box 600, Wellington, 6140, New Zealand
          \and 
              Argelander-Institut f\"{u}r Astronomie, Auf dem H\"{u}gel 71 D-53121 Bonn, Germany
          \and
               National Radio Astronomy Observatory, P.O. Box O, Socorro, NM 87801, USA
          \and
              Department of Physics and Astronomy, University of New Mexico, Albuquerque NM, 87131, USA
           }

   \date{Received; accepted}

  \abstract
  {} 
  {We study the intra-cluster magnetic field in the poor galaxy cluster Abell~194
   by complementing radio data, at different frequencies, with data in the optical and X-ray bands.}
  {We analyze new total intensity and polarization observations of Abell~194 obtained with the 
   Sardinia Radio Telescope (SRT). 
   We use the SRT data in combination with archival Very Large Array 
   observations to derive both the spectral aging and Rotation Measure (RM) 
   images of the radio galaxies 3C\,40A and 3C\,40B embedded in Abell~194.
   To obtain new additional insights in the cluster structure we investigate  
   the redshifts of 1893 galaxies, resulting in a sample of 143 fiducial cluster members.
   We analyze the available \rosat\ and \chandra\ observations to measure the electron density 
   profile of the galaxy cluster.}
   {The optical analysis indicates that Abell~194 does not
   show a major and recent cluster merger, but rather agrees with a scenario of accretion of
   small groups, mainly along the NE-SW direction. Under the minimum energy assumption, the lifetimes of synchrotron electrons 
   in 3C40\,B measured from the spectral break
   are found to be 157$\pm$11 Myrs. The break frequency image and the electron density profile inferred from the X-ray emission
   are used in combination with the RM data to constrain the intra-cluster magnetic field power spectrum.
   By assuming a Kolmogorov power law power spectrum with a minimum scale of fluctuations 
   of $\mathrm{\Lambda_{min}=1\,kpc}$, we find that the RM data in Abell~194 are well described by a 
   magnetic field with a maximum scale of fluctuations of $\mathrm{\Lambda_{max}=(64\pm24)\,kpc}$.
   We find a central magnetic field strength of $\mathrm{\langle B_0\rangle=(1.5 \pm 0.2)\,\mu G}$, the lowest ever measured so far in galaxy clusters based on Faraday rotation analysis. Further out, the field decreases with the radius following the gas density to the power of $\eta$=1.1$\pm$0.2. Comparing Abell~194 with a small sample of galaxy clusters, there is a hint of a trend between central electron densities and magnetic field strengths.}
   {}

   \titlerunning{Sardinia Radio Telescope observations of Abell~194}
   \authorrunning{F. Govoni et al.}
   \maketitle   
%

\section{Introduction}

Galaxy clusters are unique laboratories for the investigation of turbulent fluid 
motions and large-scale magnetic fields (e.g. Carilli \& Taylor 2002, 
Govoni \& Feretti 2004, Murgia 2011). 
In the last few years, several efforts have been 
focused to determine the effective strength 
and the structure of magnetic fields in galaxy clusters
and this topic represents a key project in view of the 
Square Kilometre Array (e.g. Johnston-Hollitt et al. 2015). 

Synchrotron radio halos at the center 
of galaxy clusters (e.g. Feretti et al. 2012, Ferrari et al. 2008) provide 
a direct evidence of the
presence of relativistic particles and magnetic fields associated 
with the intra-cluster medium.
In particular, the detection of polarized emission from radio halos 
is the key to investigate the magnetic field power spectrum in 
galaxy clusters 
(Murgia et al. 2004, Govoni et al. 2006, Vacca et al. 2010, Govoni et al. 2013,
Govoni et al. 2015). 
However, detecting this polarized signal is a very hard task with 
current radio facilities and so far only three examples of large-scale filamentary polarized structures possibly associated
with halo emission have been detected (A2255; Govoni et al. 2005, Pizzo et al. 2011, MACS J0717.5+3745; Bonafede et al. 2009, 
A523; Girardi et al. 2016).

Highly polarized elongated radio sources named relics are also observed at the
periphery of merging systems (e.g. Clarke \& Ensslin 2006, Bonafede et al. 2009b, van Weeren et al. 2010). 
They trace the regions where the propagation of mildly supersonic shock waves
compresses the turbulent intra-cluster magnetic field  enhancing the polarized emission and accelerating 
the relativistic electrons responsible for the synchrotron emission.

A complementary set of information on galaxy cluster magnetic fields
can be obtained from high quality RM images of powerful and 
extended radio galaxies.
The presence of a magnetized plasma between an observer and a radio source 
changes the properties of the incoming polarized emission. 
In particular, the position angle of the linearly polarized radiation 
rotates by an amount that is proportional to the line integral of the 
magnetic field along the line-of-sight times the electron density of 
the intervening medium, i.e. the so-called Faraday rotation effect. 
Therefore, information on the intra-cluster magnetic fields can be obtained, 
in conjunction with X-ray observations of the hot gas, through the analysis 
of the RM of radio galaxies in the background or in the galaxy clusters 
themselves. 
RM studies have been performed on 
statistical samples  (e.g. Clarke et al. 2001, Johnston-Hollitt \& Ekers 2004, 
Govoni et al. 2010) as well as individual 
clusters (e.g. Perley \& Taylor 1991, Taylor \& Perley 1993, Feretti et al. 1995, Feretti et al. 1999, Taylor et al. 2001, Eilek \& Owen 2002, 
Pratley et al. 2013). 
These studies reveal that magnetic fields are widespread in the intra-cluster medium, 
regardless of the presence of a diffuse radio halo emission. 

The RM distribution seen over extended radio galaxies is 
generally patchy, indicating that magnetic fields are not regularly ordered 
on cluster scales, but instead they have turbulent structures down to linear scales 
as small as a few kpc or less. Therefore, RM measurements probe 
the complex topology of the cluster magnetic field and indeed
state-of-the-art software tools and approaches based on a Fourier domain formulation have been 
developed to constrain the magnetic field power spectrum 
parameters on the basis of the RM 
images (En{\ss}lin \& Vogt 2003, Murgia et al., 2004, 
Laing et al. 2008, Kuchar \& En{\ss}lin 2011, Bonafede et al. 2013).
In some galaxy clusters and galaxy groups containing radio sources with very 
detailed RM images, the magnetic field power spectrum has been estimated
(e.g. Vogt \& En{\ss}lin 2003, Murgia et al. 2004, Vogt \& En{\ss}lin 2005, 
Govoni et al. 2006, Guidetti et al. 2008, Laing et al. 2008, 
Guidetti et al. 2010, Bonafede et al. 2010, Vacca et al. 2012).
RM data are usually consistent with volume averaged magnetic 
fields of $\simeq$0.1-1 $\mu$G over 1Mpc$^3$.
The central magnetic field strengths are typically 
a few $\mu$G, but stronger fields, with values exceeding
$\simeq$10\,$\mu$G, are measured in the inner regions of relaxed 
cooling core clusters.
There are several indications that the magnetic field intensity decreases 
going from the centre to the periphery following the cluster gas density profile. 
This has been illustrated by magneto-hydrodynamical simulations 
(see e.g. Dolag et al. 2002, Br{\"u}ggen et al. 2005, Xu et al. 2012,
Vazza et al. 2014) and confirmed in RM data. 

In this paper we aim at improving our knowledge of the intra-cluster 
magnetic field in Abell~194. This nearby (z=0.018; Struble \& Rood 1999) and 
poor (richness class $R=0$; Abell et al. 1989) galaxy cluster  
belongs to the SRT Multi-frequency Observations of Galaxy clusters (SMOG) 
sample, an early science program of the new SRT radio telescope. For our purpose, we investigated the 
total intensity and the 
polarization properties of two extended radio galaxies embedded in Abell~194,
in combination with data in optical and X-ray bands.
The paper is organized as follows.
In Sect. 2, we describe the SMOG project. 
In Sect. 3, we present the radio observations and the data reduction.
In Sect. 4, we present the radio, optical and X-ray properties of Abell~194.
In Sect. 5, we complement the SRT total intensity data with archival VLA observations to derive spectral 
aging information of the radio galaxies.
In Sect. 6, we complement the SRT polarization data with archival VLA observations to derive 
detailed multi-resolution RM images. 
In Sect. 7, we use numerical simulations to investigate the cluster magnetic field, by analyzing the RM
and polarization data.
Finally, in Sect. 8 we summarize our conclusions.

Throughout this paper we assume a $\mathrm{\Lambda CDM}$ cosmology with
$\mathrm{H_0= 71 km\, s^{-1}\, Mpc^{-1}}$,
$\mathrm{\Omega_m=0.27}$, and $\mathrm{\Omega_{\Lambda}=0.73}$. At the distance of Abell~194, 
1\arcsec\, corresponds to 0.36\,kpc.

\section{The SRT Multi-frequency Observations of Galaxy Clusters (SMOG)}

\begin{table*}
\caption{Details of the SRT observations centered on Abell~194  
$\mathrm{(RA_{J2000}=01^h25^m59.9^s;DEC_{J2000}=-01}$\deg20\arcmin33\arcsec).} 
\begin{center}         
\begin{tabular}{ccccccl}     
\hline
Frequency   & Resolution & Time on Source & SRT Proj.  & Obs. Date     & OTF Mapping     & Calibrators \\	  
(MHz)       & (\arcmin)  &    (minutes)   &            &               &                 &              \\ 
\hline
6000$-$7200 & 2.9        &    32          & S0001       & 1-Feb-2016    & 1RA             &  3C\,138, 3C\,84 \\
6000$-$7200 & 2.9        &    384         & S0001       & 3-Feb-2016    & 6RA$\times$6DEC &  3C\,138, 3C\,286, 3C\,84 \\
6000$-$7200 & 2.9        &    448         & S0001       & 6-Feb-2016    & 7RA$\times$7DEC &  3C\,286, 3C\,84\\
\hline
\multicolumn{7}{l}{\scriptsize Col. 1: SRT frequency range; Col. 2: SRT resolution; Col. 3: Time on source; Col. 4: SRT project name;}\\
\multicolumn{7}{l}{\scriptsize Col. 5: Date of observation; Col. 6: Number of images on the source; Col. 7: Calibrators.}\\
\end{tabular}   
\end{center}
\label{srtobs}
\end{table*}

The SRT is a new 64-m single dish radio telescope
located north of Cagliari, Italy. 
In its first light configuration, the SRT is equipped with three receivers: 
a 7-beam K-Band receiver (18$-$26 GHz), a mono-feed C-Band
receiver (5700$-$7700 GHz), and a coaxial dual-feed 
L/P-Band receiver (305$-$410 MHz, 1300$-$1800 MHz). 

The antenna was officially opened on September 30th 2013, 
upon completion of the technical commissioning phase (Bolli et al. 2015).
The scientific commissioning 
of the SRT was carried out in the period 2012-2015 (Prandoni et al., submitted). 
At the beginning of 2016 the first call for single dish early science
programs was issued, and the observations started on February 1st, 2016.

The SMOG project is an SRT early science program (PI M. Murgia) focused
on a wide-band and wide-field single dish spectral-polarimetric study of 
a sample of galaxy clusters. 
By comparing and complementing the SRT observations with archival radio data at 
higher resolution and at different frequencies, but also with data in
optical and X-ray bands,
we want to improve our knowledge of the non-thermal components of the intra-cluster
medium on large scale. Our aim is also to understand
the interplay between these components (relativistic particles and magnetic fields) 
and the life-cycles of cluster radio galaxies
by studying both the spectral and polarization properties of 
the radio sources with the SRT (see e.g. the case of 3C\,129; Murgia et al. 2016).
For this purpose, we selected a suitable sample of nearby galaxy clusters 
known from the literature to harbor diffuse radio halos, relics, 
or extended radio galaxies. 

We included Abell~194 in the SMOG sample because it is one of the rare clusters hosting 
more than one luminous and extended radio galaxy close to the cluster center. 
In particular, it hosts the radio source 3C\,40 (PKS 0123-016)
which is indeed constituted by two radio galaxies with distorted morphologies 
(3C\,40A and 3C\,40B).
This galaxy cluster has been extensively analyzed in the literature with
radio interferometers (e.g. O'Dea \& Owen 1985, Jetha et al. 2006,
Sakelliou et al. 2008, Bogd{\'a}n et al. 2011). 
Here we present, for the first time, total intensity and polarization single-dish observations
obtained with the SRT at 6600 MHz.
The importance of mapping the radio galaxies in Abell~194 with a single dish is that,
interferometers suffer the technical problem of not measuring the total power; the so-called missing zero spacing problem.
Indeed, they filter out structures larger than the angular 
size corresponding to their shortest spacing, limiting
the synthesis imaging of extended structures. 
Single dish telescopes are optimal for recovering 
all of the emission on large angular scales, especially at high frequencies ($>$1 GHz). 
Although single dishes typically have a low resolution, the radio galaxies at the center of Abell~194 
are sufficiently extended to be well resolved with the SRT at 6600 MHz at the resolution of 2.9$'$.

The SRT data at 6600 MHz are used in combination with archival Very Large Array 
observations at lower frequencies to derive the trend of the synchrotron spectra 
along 3C\,40A and 3C\,40B. In addition, linearly polarized emission is clearly detected for both 
sources and the resulting polarization angle images are used to produce detailed
RM images at different angular resolutions. 
3C\,40B and 3C\,40A are quite extended both in angular and linear size, 
therefore they represent ideal cases for studying the RM of the 
cluster along different lines-of-sights. In addition, the close distance 
of Abell~194 permits a detailed investigation of the cluster magnetic field structure.
Following Murgia et al. (2004) we simulated Gaussian random magnetic
field models, and we compared
the observed data and the synthetic images with a 
Bayesian approach (Vogt \& En{\ss}lin 2005), in order to constrain the strength 
and structure of the magnetic 
field associated with the intra-cluster medium.
Until recently, most of the work on cluster magnetic fields has been devoted 
to rich galaxy clusters.
Little attention has been given in the literature to magnetic fields 
associated with poor galaxy clusters like Abell~194 and galaxy 
groups (see e.g. Laing 2008, Guidetti et al. 2008, Guidetti et al. 2010). Magnetic fields 
in these system deserve to be investigated in detail  
since, being more numerous, are more representative than those of rich 
clusters.

\begin{table*}
\caption{Details of VLA archival observations of Abell~194 analyzed in this paper.}
\begin{center}
\begin{tabular} {lccllll} 
\hline
Frequency    & VLA Config. & Bandwidth &     Time on Source & VLA Proj.   & Obs. Date  & Calibrators    \\
(MHz)        &             &  (MHz)    &       (minutes)    &             &            &                \\
\hline
 C-Band               &             &           &                    &             &              &     \\
 4535/4885            &  D          &  50       &        15          & AC557       &  01-Oct-2000 &  3C\,48, 3C\,138, 0122$-$003\\
\hline
 L-Band               &             &           &                    &             &              &     \\ 
 1443/1630            &  C          &  12.5     &       142          & AV102       &  02-Jun-1984 &  3C\,48, 3C\,138, 0056$-$001\\      
 1443/1630            &  D          &  25       &        82          & AV112A      &  31-Jul-1984 &  3C\,286, 0106$+$013\\
 1465/1515            &  D          &  25       &        9           & AL252       &  18-Sep-1992 &  3C\,48, 3C\,286, 0056$-$001   \\
\hline
 P-Band               &             &           &                    &             &              &     \\
 327.5/333.0          &  B          &  3.125    &        316         & AE97        &  16-Aug-1994 & 3C\,48    \\   
 327.5/333.0          &  C          &  3.125    &        40          & AE97        &  20-Nov-1994 & 3C\,48    \\   
\hline
\multicolumn{7}{l}{\scriptsize Col. 1: Observing frequency (IF1/IF2); Col. 2: VLA configuration; Col. 3: Observing bandwidth; Col. 4: Time on source;}\\
\multicolumn{7}{l}{\scriptsize Col. 5: VLA project name; Col. 6: Date of observation; Col. 7: Calibrators.}\\
\end{tabular}
\label{vla}
\end{center}
\end{table*}


\section{Radio observations and data reduction}

\subsection{SRT observations}

We observed with the SRT, for a total exposure time of about 14.4 hours,
an area of 1deg$\times$1deg~ centered on the galaxy 
cluster Abell~194 using the C-Band receiver.
Full-Stokes parameters were acquired with the SARDARA back-end
(SArdinia Roach2-based Digital Architecture 
for Radio Astronomy; Melis et al., submitted), one of the back-ends available
at the SRT (Melis et al. 2014). We used the correlator
configuration with 1500 MHz and 1024 frequency channels of approximately 1.46\,MHz in width. We observed in the frequency range 6000$-$7200 MHz, at a central frequency of 6600 MHz.
We performed several on-the-fly (OTF) mappings in the equatorial frame 
alternating the right ascension (RA) and declination (DEC). 
The telescope scanning speed was set to 6 arcmin/s and the scans were 
separated
by 0.7\arcmin~to properly sample the SRT beam whose full width at half maximum (FWHM) is 2.9\arcmin~in this frequency range. 
We recorded the data stream sampling at 33 spectra 
per second, therefore individual samples were separated on the sky 
by 10.9\arcsec~along the scanning direction.
A summary of the SRT observations is listed in Table\,\ref{srtobs}.
Data reduction was performed with the proprietary Single-dish Spectral-polarimetry 
Software (SCUBE; Murgia et al. 2016).

Band-pass, flux density, and polarization calibration were done
with at least four cross scans on source calibrators performed at the beginning and 
at the end of each observing section.

Band-pass and flux density calibration were performed by observing 3C\,286 and 3C\,138,  
assuming the flux density scale of Perley \& Butler (2013).
After a first bandpass and flux density calibration cycle, persistent radio frequency interference (RFI) were flagged 
using observations of a cold part of the sky.
The flagged data were then used to repeat the bandpass and the flux density calibration
for a finer RFI flagging.
The procedure was iterated a few times, eliminating all of the most obvious RFI.
We applied the gain-elevation curve correction to account for the gain variation  
with elevation due to the telescope structure gravitational stress change. 

We performed the polarization calibration by correcting the instrumental polarization
and the absolute polarization angle.
The on-axis instrumental polarization was determined through observations of the bright unpolarized source 3C\,84. 
The leakage of Stokes I into Q and U is in general less than 2\% across the band, 
with a rms scatter of $0.7 - 0.8$\%.
We fixed the absolute position of the polarization angle using as reference the primary 
polarization calibrator 3C\,286 and 3C\,138.
The difference between the observed and predicted position angle according to Perley \& Butler (2013) 
was determined, and corrected channel-by-channel.
The calibrated position angle was within the expected value 
of 33\deg\, and $-$11.9\deg\, for 3C\,286 and 3C\,138, respectively, with a rms scatter of $\pm 1$\deg.

In the following we describe the total intensity and the polarization imaging.
For further details on the calibration and data handling of the
SRT observations see Murgia et al. (2016).

\subsubsection{Total intensity imaging}

The imaging was performed in SCUBE by subtracting the baseline from the calibrated telescope 
scans and by  projecting the data in a regular three-dimensional grid.
At 6600\, MHz we used a spatial resolution of 0.7\arcmin/pixel (corresponding to the separation of the telescope scans),
which is enough in our case to sample the beam FWHM with four pixels.

As a first step, the baseline was subtracted 
by a linear fit involving only the 10\% of data at the beginning and the end 
of each scan. 
The baseline removal was performed channel-by-channel for each scan.
All frequency cubes obtained by gridding the scans along the two orthogonal axes (RA and DEC) were then stacked together to produce full-Stokes I, Q, U images 
of an area of 1 square degree centered on the galaxy cluster Abell~194. In the combination, the individual image cubes 
were averaged and de-stripped by mixing their Stationary Wavelet Transform (SWT) coefficients (see Murgia et al. 2016, for details). 
We then used the higher Signal-to-Noise (S/N) image cubes obtained from the SWT stacking as a prior model to refine the baseline fit. 
The baseline subtraction procedure was then repeated including not just the 10\% from the begin and the end of each scan. 
In the refinement stage, the baseline was represented with a 2nd order polynomial using the regions of the scans free of radio sources.
A new SWT stacking was then performed and the process was iterated a few more times until the convergence was reached.

Close to the cluster center, the dynamical range of the C-Band total intensity image was limited by the telescope beam 
pattern rather than by the image sensitivity. 
We used in SCUBE a beam model pattern (Murgia et al. 2016) for a proper deconvolution of the 
sky image from the antenna pattern.
The deconvolution algorithm interactively finds the peak in the image obtained from the SWT stacking of all images, 
and subtracts a fixed gain fraction (typically 0.1) of this point 
source flux convolved with the re-projected telescope ``dirty beam model'' from the individual images. 
In the re-projection, the exact elevation and parallactic angle for
each pixel in the unstacked images are used. The residual images were
stacked again and the CLEAN continued until a threshold condition was reached.
Given the low level of the beam side lobes compared to interferometric images, 
a shallow deconvolution was sufficient in our case, and we decided to stop 
the CLEAN at the first negative component encountered. As a final step, CLEAN components at the same position were merged, 
smoothed with a circular Gaussian beam with FWHM 2.9\arcmin, and then 
restored back in the residuals image to obtain a CLEANed image.

\subsubsection{Polarization imaging}

The polarization imaging at C-Band of Stokes parameters Q and U was performed following the same procedures described for the total intensity imaging: baseline subtraction, 
gridding, and SWT stacking. There were no critical dynamic range issues with the polarization image, and thus no deconvolution was
required. However, since the contribution of the off-axis instrumental polarization can affect the quality of polarization data if bright sources are 
present in the image, we corrected for the off-axis instrumental polarization by deconvolving the Stokes Q and U beam patterns.
In particular, SCUBE uses the CLEAN components derived from the deconvolution of the beam pattern from the total intensity image to subtract 
the spurious off-axis polarization from each individual Q and U scans before their stacking. 
The off-axis instrumental polarization level compared to the Stokes I peak is 0.3\%. 

Finally, polarized intensity $\mathrm{P=\sqrt{Q^2+U^2}}$ (corrected for the positive bias), 
fractional polarization $\mathrm{FPOL=P/I}$ and position angle of
polarization $\mathrm{\Psi=0.5\tan^{-1}(U/Q)}$ images were then derived from the I, Q, 
and U images.

\subsection{Archival VLA observations}

We analyzed archival observations obtained with the VLA at different frequencies and configurations.
The details of the observations are provided in Table\,\ref{vla}.
The data were reduced using the NRAO's Astronomical Image Processing
System (AIPS) package. 

The data in C-Band and in L-Band were calibrated in amplitude, phase, and
polarization angle using the standard calibration procedure. 
The flux density of the calibrators have been calculated accordingly to the
flux density scale of Perley \& Butler (2013).
Phase calibration was derived from nearby sources, periodically observed over a wide range in parallactic angle.
The radio sources 3C\,286 or 3C\,138 were used as reference for the absolute
polarization angles. Phase calibrators were also used 
to separate the source polarization properties from the antenna polarizations.  
Imaging was performed following the standard procedure: Fourier-Transform, Clean and Restore. 
A few cycles of self-calibration were applied 
when they have proven to be useful to remove residual phase variations.
Images of the Stokes parameters I, Q, and U, have been produced for 
each frequency and configuration separately providing different angular resolutions. 
Since the C-Band data consist of three separate short pointings, in this case, 
the final I, Q, and U images were obtained by mosaicing the three
different pointings with the AIPS task FLATN.
Finally, P, FPOL, and $\Psi$ images were then derived from the I, Q, and U images.
 
Data in P-Band were obtained in spectral line mode dividing the bandwidth of 3.125 MHz 
in 31 channels.
A data editing was made in order to excise RFI channel by channel. We performed the amplitude and bandpass calibration with 
the source 3C\,48. The flux density of the calibrator was calculated accordingly to the
low frequency coefficients of Scaife \& Heald (2012).
In the imaging, the data were averaged into five channels. The data were mapped using a wide-field imaging 
technique, which corrects for 
distortions in the image caused by the non-coplanarity of the VLA over a wide field of view. 
A set of small overlapping maps was used to cover the central area of about $\sim$2.4\deg\, in radius 
(Cornwell \& Perley 1992). 
However, at this frequency confusion lobes of sources far from the center of the field are still present. Thus, we also obtained images of 
strong sources in an area of about $\sim$6\deg\, in radius, searched in the NRAO VLA Sky Survey (NVSS; Condon et al. 1998) catalog. 
All these "facets'' were included in CLEAN and used for several loops of phase self-calibration (Perley 1999). 
To improve the u-v coverage and sensitivity we combined the data sets in B and C configuration. 

\section{Multi-wavelengths properties of Abell~194}

\begin{figure*}
\centering
\includegraphics[width=18 cm]{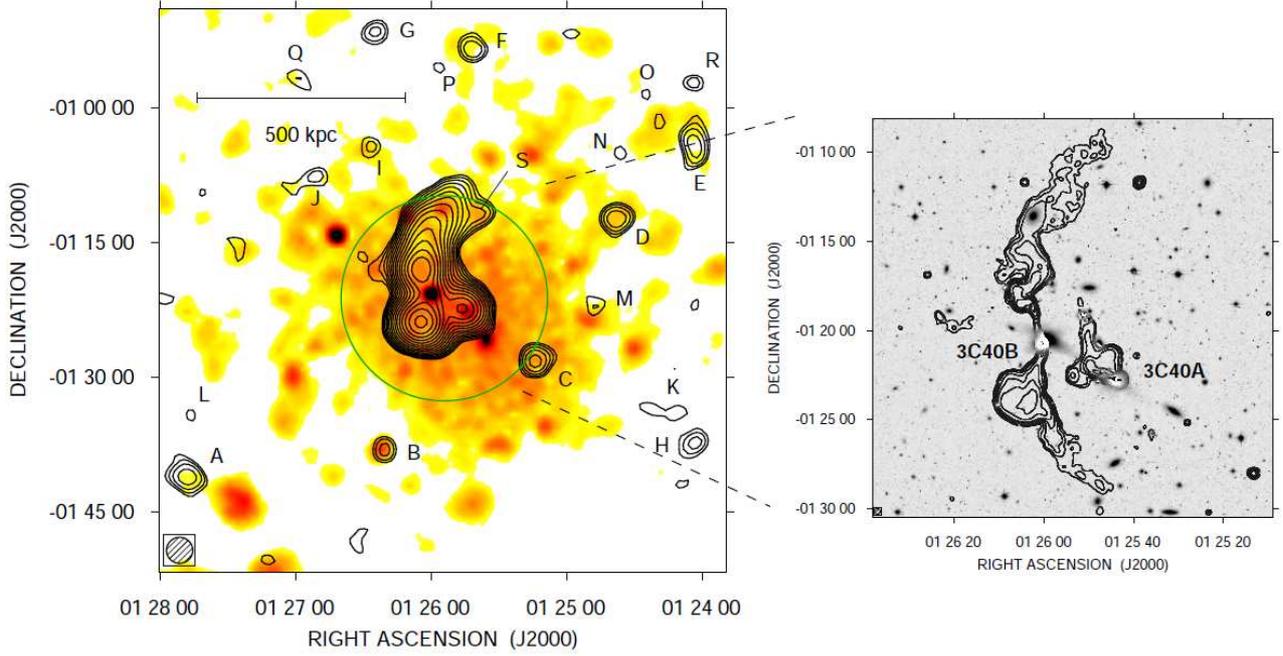}
\caption[]{
{\it Left}: SRT radio image at 6600 MHz (contours) overlaid on the \rosat\ PSPC X-ray image (colors) in 
the $0.4-2$\,keV band. The SRT image results from the spectral average of the
bandwidth between 6000 and 7200 MHz. It has a sensitivity of 1 mJy/beam 
and an angular resolution of 2.9\arcmin. The first radio contour is drawn at 3 mJy/beam 
$\mathrm{(3\sigma_I)}$ and the rest are spaced by a factor of $\sqrt{2}$.
SRT sources with an NVSS counterpart are labeled with the letters A to S. 
The X-ray image (ID=800316p) is exposure corrected $\mathrm{(T_{exp}\simeq24 ksec)}$ and
adaptively smoothed (see Sect.\,\ref{xray}).
The green circle is centered on the cluster X-ray centroid
and the radius indicates the cluster core radius $r_{\rm c}$.  
{\it Right}: VLA radio image at 1443 MHz (contours) overlaid on the optical emission in the $g^{\rm Mega}$ band
(grey-scale). The VLA image has a sensitivity of 0.34 mJy/beam 
and an angular resolution of 19\arcsec.
The first radio contour is drawn at 1 mJy/beam $\mathrm{(3\sigma_I)}$ and the 
rest are spaced by a factor of 2.
}
\label{Fig_OtticoX}
\end{figure*}

In the following we present the radio, optical, 
and X-ray properties of the galaxy cluster Abell~194.

\subsection{Radio properties}
\label{radio}

In Fig.\,\ref{Fig_OtticoX}, we show the radio, optical, and X-ray emission of Abell~194.
The field of view of the left panel of Fig.\,\ref{Fig_OtticoX} is $\simeq$1.3$\times$1.3 Mpc$^2$.
In this panel, the contours of the CLEANed radio image obtained with the SRT at 6600 MHz
are overlaid on the X-ray \rosat\ PSPC image in the $0.4-2$\,keV band
(see Sect.\,\ref{xray}).
The SRT image was obtained by averaging all the frequency channels from 6000 MHz
to 7200 MHz. We reached a final noise level of 1 mJy/beam 
and an angular resolution of 2.9\arcmin~FWHM. 
The radio galaxy 3C\,40B, close to the cluster X-ray center,
extends for about 20\arcmin. 
The peak brightness of 3C\,40B ($\simeq$600 mJy/beam) is located in the 
southern lobe. The narrow angle-tail radio galaxy 3C\,40A  
(peak brightness $\simeq$150 mJy/beam), is only slightly resolved 
at the SRT resolution and it is not clearly separated from 3C\,40B.

The details of the morphology of the two radio galaxies can be appreciated
in the right panel of Fig.\,\ref{Fig_OtticoX}, where we show a 
field of view of $\simeq$0.5$\times$0.5 Mpc$^2$. In this panel, 
the contours 
of the radio image obtained with the VLA at 1443 MHz are overlaid 
on the optical emission of the cluster.
We retrieved the optical image in the $g^{\rm Mega}$ band 
from the CADC 
Megapipe\footnote{http://www.cadc-ccda.hia-iha.nrc-cnrc.gc.ca/en/megapipe/} archive (Gwyn 2008).
The VLA radio image was obtained from the L-Band data set in C configuration.
It has a sensitivity of 0.34 mJy/beam and an angular resolution of 19\arcsec.
The distorted morphology of the two radio galaxies 3C\,40B and 3C\,40A is well visible.
Both radio galaxies show an optical counterpart and their host galaxies are separated 
by 4.6\arcmin~($\simeq$ 100 kpc). The core of the extended source 3C\,40B is associated with NGC\,547, which is known to  
form a dumbbell system with the galaxy NGC\,545 (e.g. Fasano et al. 1996). 
3C\,40A is a narrow angle-tail radio galaxy associated 
with the galaxy NGC\,541 (O'Dea \& Owen 1985).
The jet emanating from 3C\,40A is believed to be responsible 
for triggering star formation in Minkowski's object (e.g. van Breugel
et al. 1985, Brodie et al. 1985, Croft et al. 2006), a 
star-forming peculiar galaxy near NGC\,541.  

\begin{table*}
\caption{Flux density measurements of faint radio sources detected in the Abell~194 field.}             
\begin{center}
\begin{tabular}{lcccccc}     
\hline
Source         &  Label     & RA J2000       & DEC J2000                & $\mathrm{S_{6600 MHz}}$        &  $\mathrm{S_{1400 MHz}}$   & $\mathrm{\alpha_{\rm 6600\,MHz}^{\rm 1400\,MHz}}$   \\	  
               &            & (h m s)        &  (d \arcmin~\arcsec)     & (mJy)               &  (mJy)          &                                     \\
\hline
NVSSJ012748-014134  &  A         &  01 27 48      &   $-$01 41 05     & $ 11.5 \pm 0.7   $    & $24.6 \pm  2.3 $   & $ 0.49\pm0.13 $           \\ 
NVSSJ012622-013756  &  B         &  01 26 21      &   $-$01 38 00     & $  5.3 \pm 0.7   $    & $11.0 \pm  0.6 $   & $ 0.47\pm0.29 $           \\ 
NVSSJ012513-012800  &  C         &  01 25 13      &   $-$01 27 58     & $ 13.3 \pm 0.8   $    & $32.9 \pm  1.1 $   & $ 0.58\pm0.10 $           \\ 
NVSSJ012438-011232  &  D         &  01 24 38      &   $-$01 12 25     & $ 11.0 \pm 0.8   $    & $35.7 \pm  1.5 $   & $ 0.76\pm0.10 $           \\ 
Blends of NVSS sources &  E      &  01 24 03      &   $-$01 04 11     & $ 13.9 \pm 1.2   $    &      $-$             &   $-$                     \\ 
NVSSJ012542-005302  &  F         &  01 25 41      &   $-$00 53 06     & $  7.7 \pm 0.7   $    & $8.4  \pm 1.4  $   & $ 0.06\pm0.36 $           \\ 
NVSSJ012624-005101  &  G         &  01 26 24      &   $-$00 51 26     & $  5.8 \pm 0.8   $    & $21.0 \pm 1.1  $   & $ 0.83\pm0.18 $           \\ 
NVSSJ012401-013706  &  H         &  01 24 03      &   $-$01 37 18     & $  9.7 \pm 2.1   $    & $19.3 \pm 0.8  $   & $ 0.44\pm0.17 $           \\ 
NVSSJ012628-010418  &  I         &  01 26 26      &   $-$01 04 19     & $  4.1 \pm 0.8   $    & $19.4 \pm 1.4  $   & $ 1.00\pm0.22 $           \\ 
Blends of NVSS sources &  J      &  01 26 51      &   $-$01 07 35     & $  4.8 \pm 0.8   $    &      $-$             &   $-$                    \\ 
NVSSJ012410-013357  &  K         &  01 24 10      &   $-$01 33 51     & $  5.0 \pm 0.8   $    & $7.3 \pm  0.6  $   & $ 0.24\pm0.42 $          \\ 
NVSSJ012746-013446  &  L         &  01 27 46      &   $-$01 34 10     & $  2.7 \pm 0.7   $    & $3.9 \pm 0.5 $     & $0.24\pm0.78 $           \\ 
NVSSJ012450-012208  &  M         &  01 24 47      &   $-$01 22 00     & $  3.7 \pm 0.8   $    & $ 5.1 \pm  0.7  $  & $ 0.21\pm0.60 $                 \\ 
NVSSJ012431-010459  &  N         &  01 24 36      &   $-$01 04 46     & $  2.9 \pm 0.7   $    & $ 7.1 \pm  1.5  $  & $ 0.58\pm0.47 $                 \\ 
NVSSJ012426-005827  &  O         &  01 24 25      &   $-$00 58 25     & $  4.0 \pm 1.6   $    & $ 7.9 \pm  0.6  $  & $ 0.44\pm0.40 $                 \\ 
NVSSJ012557-005442  &  P         &  01 25 57      &   $-$00 55 19     & $  2.9 \pm 0.6   $    & $ 4.6 \pm 0.7  $   & $ 0.30\pm0.67 $                  \\ 
Blends of NVSS sources &  Q      &  01 26 58      &   $-$00 56 46     & $  4.6 \pm 0.7   $    &      $-$           &     $-$                                \\
NVSSJ012406-005638  &  R         &  01 24 03      &   $-$00 57 05     & $  3.6 \pm 0.7  $     & $4.0  \pm 0.5$     & $ 0.07\pm0.75 $                          \\ 
NVSSJ012538-011139  &  S         &     $-$        &      $-$          &     $-$               & $16.1 \pm 1.5$     &     $-$                        \\ 
\hline 
\multicolumn{7}{l}{\scriptsize Col. 1: NVSS cross-identification; Col. 2: Source label (see Fig.\,\ref{Fig_OtticoX}); Cols. 3 and 4: 
Coordinates of the peak intensity in the SRT image;}\\
\multicolumn{7}{l}{\scriptsize Col. 5: Flux density at 6600\,MHz, taken from the SRT image; Col. 6: Flux density at 1400\,MHz, taken from the NVSS;}\\
\multicolumn{7}{l}{\scriptsize Col. 7: Spectral index between 1400 and 6600\,MHz.}\\
\end{tabular}
\label{identificazione}
\end{center} 
\end{table*} 

In addition to 3C\,40A and 3C\,40B, some other faint radio sources 
have been detected in the Abell~194 field. In 
the left panel of Fig.\,\ref{Fig_OtticoX}, SRT sources with an NVSS 
counterpart are labeled with the letters A to S. 
Sources labeled with E, J, and Q are actually blends 
of multiple NVSS sources. There are also a few sources
visible in the NVSS but not detected at the sensitivity level of the SRT image. 
These are likely steep spectral index radio sources. 
The SRT contours show an elongation in the northern lobe of 3C\,40B toward 
west. This elongation is likely due to the point source labelled with S, 
which is detected both in the 1443 MHz VLA image at 19\arcsec resolution  
in the right panel of Fig.\,\ref{Fig_OtticoX}, and in the NVSS. 
The 1443 MHz image also shows the presence of another point source located 
on the east of the northern lobe of 3C\,40B. 
This point source is blended with 3C\,40B both at the SRT and at 
the NVSS resolution.

Given that 3C\,40A and 3C\,40B are not clearly separated at the SRT resolution, 
we calculated the flux density for the two sources together, by integrating 
the total intensity image at 6600 MHz down to the $\mathrm{3\sigma_I}$ isophote. 
It results $\simeq$1.72$\pm$0.05 Jy. 
This flux contains also the flux of the two discrete sources located in 
the northern lobe of 3C\,40B mentioned above.

In Table\,\ref{identificazione}, we list the basic properties of the
faint radio sources detected in the field of the SRT image.
For the unresolved sources, we calculated the flux density by 
means of a bi-dimensional Gaussian fit.
Along with the SRT coordinates and the SRT flux densities, 
we also report their NVSS name, the NVSS flux density
at 1400 MHz, and the global spectral indices ($S_{\nu}\propto \nu^{-\alpha}$) 
between 1400 and 6600\,MHz.

\begin{figure*}
\centering
\includegraphics[width=16 cm]{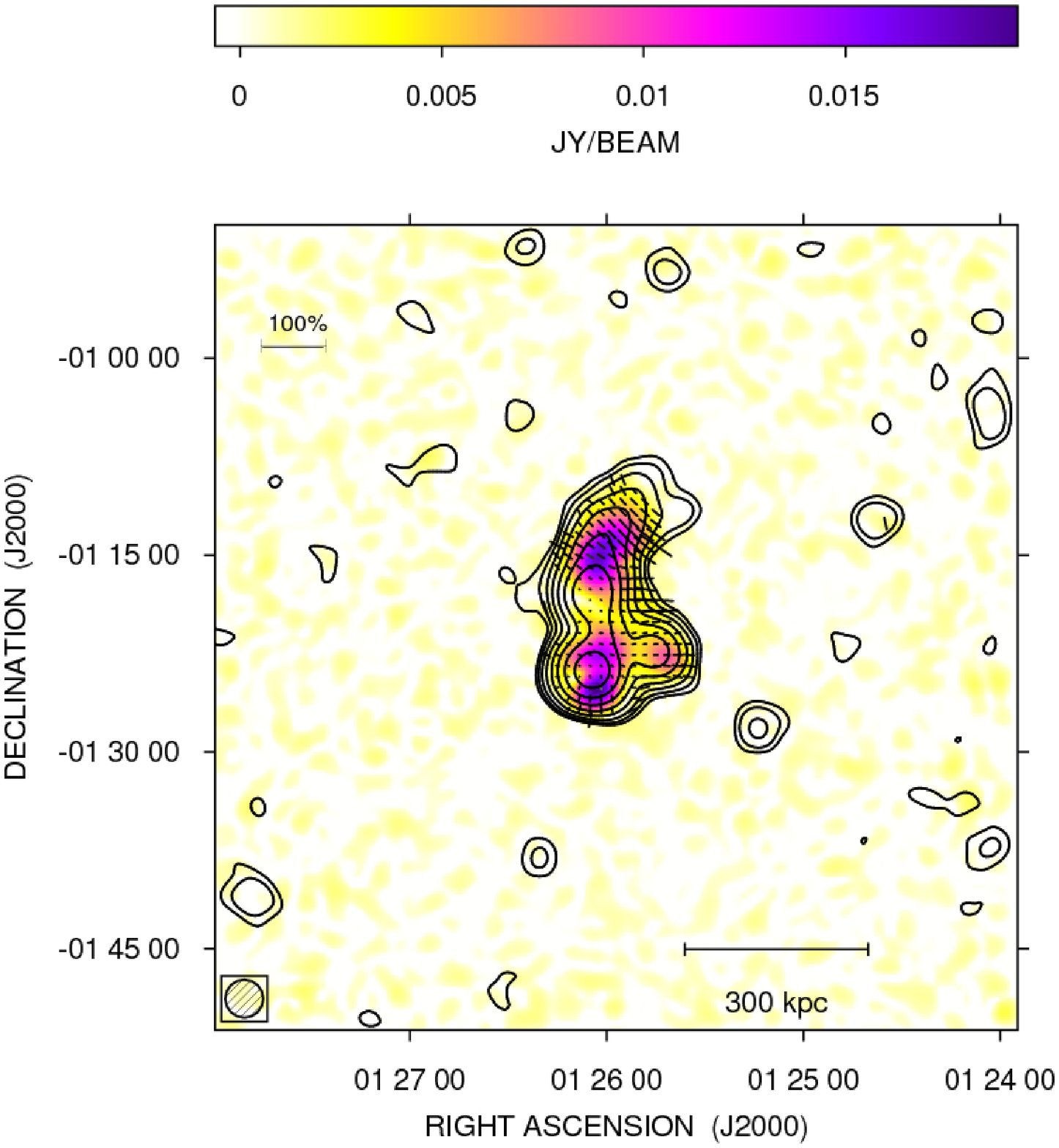}
\caption[]{
SRT linearly-polarized intensity image at 6600 MHz of the galaxy cluster Abell~194, 
resulting from the spectral average of the bandwidth between 6000 and 7200 MHz. 
The noise level, after the correction for the polarization bias, is $\mathrm{\sigma_P}$=0.5 mJy/beam. 
The FWHM beam is 2.9\arcmin, as indicated in the bottom-left corner. Contours refer to the
total intensity image. Levels start at 3 mJy/beam $\mathrm{(3\sigma_I)}$ and increase by a factor of 2.
The electric field (E-field) polarization
vectors are traced only for those pixels where the total intensity signal is above $\mathrm{5\sigma_I}$,  
the error on the polarization angle is less than 10\deg, and the fractional polarization is above $\mathrm{3\sigma_{FPOL}}$. 
The length of the vectors is proportional to the polarization percentage (with 100\% represented by the bar in the top-left corner).
}
\label{Fig_SRT_Pol}
\end{figure*}

In Fig.\,\ref{Fig_SRT_Pol}, we show the total intensity SRT contours 
levels overlaid
on the linearly-polarized intensity image P 
at 6600 MHz.
The polarized intensity was corrected for both the on-axis and the off-axis instrumental polarization. 
The noise level, after the correction for the polarization bias, is $\mathrm{\sigma_{\rm P}}$=0.5 mJy/beam. 
Polarization is detected for both 3C\,40B and 3C\,40A with a global fractional polarization of $\simeq$9\%.
The peak polarized intensity is of $\simeq$19 mJy/beam, located in the 
south lobe of 3C\,40B, not matching the total intensity peak.
The origin of this mismatch could be attributed to different effects. A possibility
is that the magnetic field inside the radio source's lobe is not completely ordered. Indeed, along the line-of-sight
at the position of the peak intensity, we may have by chance two (or more) magnetic field structures not perfectly 
aligned in the 
plane of the sky so that the polarized intensity is reduced, while the total intensity is
unaffected. This depolarization is a pure geometrical effect 
related to the intrinsic ordering of the source's magnetic field and may be present even if there is no Faraday 
rotation inside and/or outside the radio emitting plasma. Another possibility is that Faraday Rotation
is occurring in an external screen and the peak intensity is located in projection in a region of a high RM
gradient (see RM image in Fig.\,\ref{rm_srt}). 
In  this case, the beam depolarization is expected to reduce the polarized signal but not the total 
intensity. Finally, there could be also internal Faraday 
Rotation, but the presence of an X-ray cavity (see Sect.\,\ref{xray}), suggests that the southern lobe
is devoided of thermal gas and furthermore, the observed trend of the
polarized angles are consistent with the $\lambda^2$-law that points in favour of an external 
Faraday screen (see Sect.\,\ref{rotmes}).

In 3C\,40B the fractional polarization increase along the source from 1-3\% in the central brighter part up to
15-18\% in the low surface brightness associated with the northern lobe.
The other faint sources detected in the Abell~194 field are not significantly polarized in the SRT image.

\subsection{Optical properties}
\label{optical}

\begin{figure}
\centering
\includegraphics[width=8.5 cm]{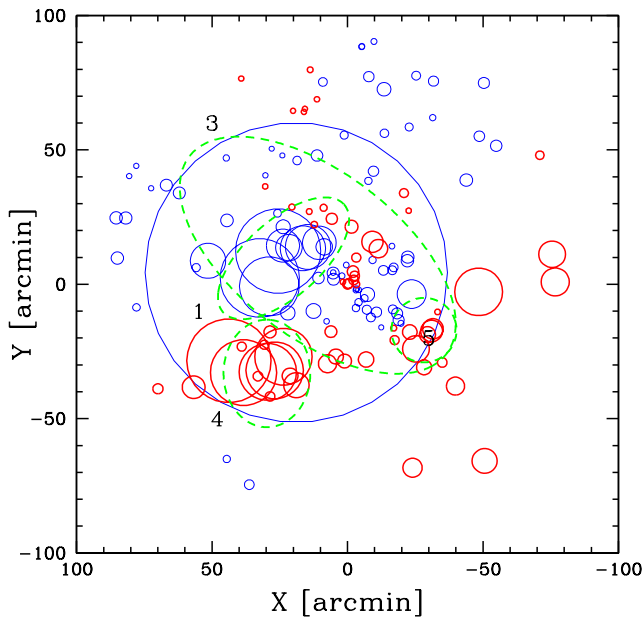}
\caption[]{Spatial
distribution of the 143 cluster member galaxies.  The larger the
circle, the larger is the deviation of the local mean velocity, as
computed on the galaxy and its ten neighbors, from the global mean
velocity. Blue/thin-line and red/thick-line circles show where the
local value of mean velocity is smaller or larger than the global
value. Green dashed ellipses indicate the regions of the subgroups
detected by Nikogossyan et al. (1999, see their Fig.\,13 - the subgroup 
No.\,2 is outside of our field). The center is fixed on the NGC\,547 galaxy.}
\label{figd}

\end{figure}
 
\begin{figure} 
\centering
\includegraphics[width=8.7 cm]{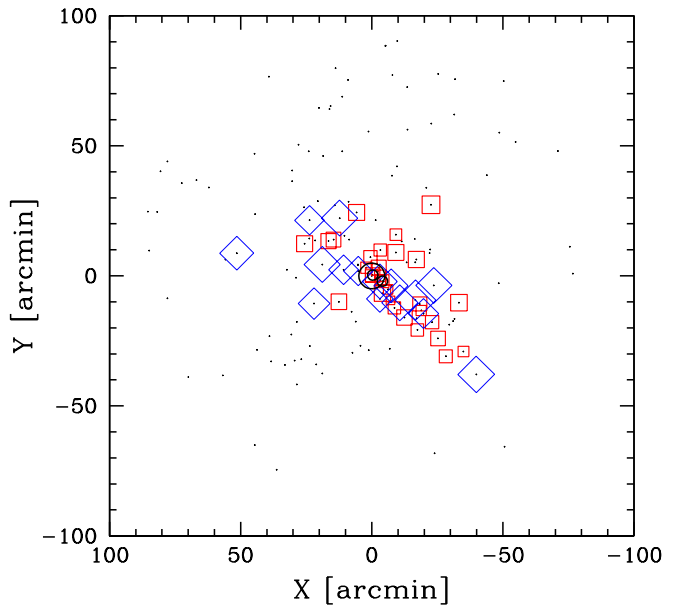}
\caption[]{Spatial
distribution of the 143 cluster members (small black points). The
galaxies of the two main subgroups detected in the 3D-DEDICA analysis
are indicated by large symbols. Red squares and blue rotated squares
indicate galaxies of the main and the secondary subgroups, respectively. The 
secondary subgroup is characterized by a lower velocity. The larger the symbol size, the
larger is the deviation of the galaxy from the mean velocity.
The center is fixed on the NGC\,547 galaxy (large black circle).  NGC\,545
and NGC\,541 are indicated by small circles.
} 
\label{figshow} 
\end{figure}

One of the intriguing properties of Abell~194 is that it appears as a ``linear cluster'', as its galaxy distribution and X-ray emission are both linearly
elongated along the NE-SW direction 
(Rood \& Sastry 1971, Struble \& Rood 1987; Chapman et al. 1988; 
Nikogossyan et al. 1999; Jones \& Forman 1999).

Previous studies of Abell~194, based on redshift data, have
found a low value of the velocity dispersion of galaxies within the
cluster ($\sim$ 350$-$400 \kss; Chapman et al. 1988, Girardi et
al. 1998) and a low value of the mass. Two independent
analyses agree in determining a mass of $M_{200}\sim 1 \times 10^{14}$
\msun within the radius\footnote{The radius $R_{\delta}$ is the radius
  of a sphere with mass overdensity $\delta$ times the critical
  density at the redshift of the galaxy system.}  $R_{200}\sim 1$ Mpc
(Girardi et al. 1998, Rines et al. 2003). Early
analyses of redshift-data samples of $\sim$ 100-150 galaxies within
3-6 Mpc, mostly derived by Chapman et al. (1988), confirmed that
Abell~194 is formed of a main system elongated along the NE-SW direction 
and detected minor substructure in both the central and external cluster
regions (Girardi et al. 1997, Barton et al. 1998,
Nikogossyan et al. 1999).

To obtain new additional insights in the cluster
structure, we considered more recent redshift data.  
Rines et al. (2003) compiled redshifts
from the literature as collected by the NASA/IPAC Extragalactic
Database (NED), including the first Sloan Digital Sky Survey (SDSS)
data.  Here, we also added further data extracted from the last SDSS
release (DR12).  In particular, to analyze
the 2$R_{200}$ cluster region, we considered galaxies within 2 Mpc
($\sim 93$\arcmin) from the cluster center (here taken as the galaxy
3C\,40B/NGC\,547).
Our galaxy sample consists of 1893 galaxies. 
After the application of the P+G membership selection
(Fadda et al. 1996; Girardi et al. 2015) we obtained a sample of 
143 fiducial members.
The P+G membership selection is a two steps method. First,
we used the 1D adaptive-kernel method DEDICA (Pisani 1993)
to detect the significant cluster peak in the velocity
distribution.  All the galaxies assigned to the cluster peak are
analyzed in the second step which uses the combination of position
and velocity information: the ``shifting gapper'' method by Fadda et
al. (1996).  This procedure rejects galaxies that are too
far in velocity from the main body of galaxies and within a fixed
bin that shifts along the distance from the cluster center.  The
procedure is iterated until the number of cluster members converges
to a stable value.

The comparison with the corresponding galaxy samples of Chapman et
al. (1988), formed of 84 galaxies (67 members), shows that we
have doubled the data sample and stresses the difficulty of improving
the Abell~194 sample when going down to fainter luminosities.  
Unlike the Chapman et al. (1988) sample, the sampling and
completeness of both our sample and that of Rines et al. (2003)
are not uniform and, in particular, since the
center of Abell~194 is at the border of a SDSS strip, the southern clusters
regions are undersampled with respect to the northern ones.  
As a consequence, we limited our analysis of substructure to the velocity
distribution and to the position-velocity combined data.

By applying the biweight estimator to the 143 cluster members (Beers
et al. 1990, ROSTAT software), we computed a mean cluster
line-of-sight (LOS) velocity $\left<V\right>=\left<cz\right>
=(5\,375\pm$143) \kss, corresponding to a mean cluster redshift
$\left<z\right>=0.017929\pm0.0004$ and a LOS velocity dispersion
$\sigma_V=425_{-30}^{+34}$ \kss, in good agreement with the estimates
by Rines et al. (2003, see their Table\,2). There is no
evidence of non Gaussianity in the galaxy velocity distribution
according to two robust shape estimators, the asymmetry index and the
tail index, and the scaled tail index (Bird \& Beers 1993).
We also verified that the three luminous galaxies in the cluster core
(NGC\,547, NGC\,545, and NGC\,541) have no evidence of peculiar velocity
according to the indicator test by Gebhardt \& Beers (1991).

We applied the $\Delta$-statistics devised by Dressler \& Schectman
(1988, hereafter DS-test), which is a powerful test for 3D
substructure. The significance is based on 1000 Monte Carlo simulated
clusters obtained shuffling galaxies velocities with respect to their
positions. We detected a marginal evidence of substructure (at the
91.3\% confidence level). Fig.\,\ref{figd} shows the comparison of the DS
bubble-plot, here obtained considering only the local velocity
kinematical DS indicator, with the subgroups detected by Nikogossyan
et al. (1999). The most important subgroup detected by
Nikogossyan et al. (1999), the No.~3, traces the NE-SW
elongated structure. Inside this, the regions characterized by low or
high local velocity correspond to their No.~1 and No.~5 subgroups. A
SE region characterized by a high local velocity corresponds to their
No.~4 subgroup, the only one outside the main NE-SW cluster structure.
The above agreement is particularly meaningful when considering that
the Nikogossyan et al. (1999) and our results are based on
quite different samples and analyses. In particular, their
hierarchical-tree analysis weights galaxies with their luminosity,
while no weight is applied in our DS-test and plot.
The galaxy with a very large blue/thin circle has a difference from
the mean cz of 488 km/s and lies at 0.42 Mpc from the cluster center,
that is well inside the caustic lines reported by Rines et al. (2003,
see their Fig.2) and thus it is definitely a cluster member. However,
this galaxy lies at the center of a region inhabited by several
galaxies at low velocity, resulting in the large size of the circle.
In fact, as mentioned in the caption, the circle size refers to the local mean velocity 
as computed with respect to the galaxy and its ten neighbors.

We also performed
the 3D optimized adaptive-kernel method of Pisani (1993, 1996; 
hereafter 3D-DEDICA, see also Girardi et al. 2016). The method detects two important density peaks,
significant at the $>99.99\%$ confidence level of 31 and 15 galaxies.  Minor
subgroups have very low density and/or richness and are not discussed.
Fig.\,\ref{figshow} shows as both the two DEDICA subgroups are
strongly elongated and trace the NE-SW direction but have different
velocities. The main one has a velocity peak of 5\,463 \kss, close to
the mean cluster velocity, and contains NGC\,547, NGC\,545, and NGC\,541.
The secondary one has a lower velocity (4\,897 \kss).

The picture resulting from new and previous optical results agree in that Abell~194 does not
show trace of a major and recent cluster merger (e.g., as in the case
of a bimodal head-on merger), but rather agrees with a scenario of accretion of
small groups, mainly along the NE-SW axis.

\subsection{X-ray properties}
\label{xray}

\begin{figure} 
\centering
\includegraphics[width=8.5 cm]{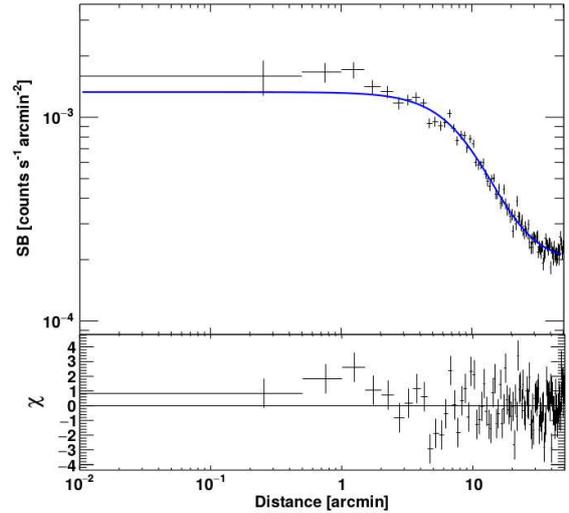}
\caption[]{
Surface brightness profile of the X-ray emission of
Abell~194. The profile is centered at the position of the X-ray centroid. 
The best fit $\beta$-model is shown in blue.
}
\label{betamodel}
\end{figure}

The cluster Abell~194 has been observed in X-rays with the ASCA, 
\rosat, \chandra, and XMM satellites.
Sakelliou et al. (2008), investigated
the cluster relying on XMM and radio observations.
The X-ray data do not show any signs of features expected from
recent cluster merger activity.
They concluded that the central region of Abell~194 is relatively 
quiescent and is not suffering a major merger event, in agreement with the optical analysis
described in Sect.\,\ref{optical}.
Bogd{\'a}n et al. (2011), analyzed X-ray observations with \chandra\ 
and \rosat\ satellites. They mapped 
the dynamics of the galaxy cluster and also detected a large 
X-ray cavity formed by the southern radio lobe arising from 3C\,40B.
Therefore, this target is particularly interesting for 
Faraday rotation studies because the presence of an X-ray cavity 
indicates that the rotation of the polarization plane is likely to 
occur entirely in the intra-cluster medium, since comparatively little
thermal gas should be present inside the radio-emitting plasma.

The temperature profile and the cooling time of 21.6$\pm$4.41 Gyr
determined by Lovisari et al. (2015), and the temperature map by Bogd{\'a}n et al. (2011), 
indicate that Abell~194 does not harbor a cool core.

We analyzed the \rosat\ PSPC pointed observation of Abell~194 following the
same procedure of Eckert et al. (2012) to which we refer for a more detailed
description. Here we briefly mention the main steps of the reduction and
of the analysis. The \rosat\ Extended Source Analysis Software
(Snowden et al. 1994) was used for the data reduction. The contribution of the
various background components such as scattered solar X-rays, long term
enhancements and particle background has been taken into account and combined
to get a map of all the non-cosmic background components to be subtracted. 
We then extracted the image in the R37 \rosat\ energy band (0.4-2 keV) 
corrected for vignetting effect with the corresponding exposure map. Point 
sources have been detected and excluded up to a constant flux threshold in 
order to ensure a constant resolved fraction of the Cosmic X-ray Background (CXB) 
over the field of view. The surface brightness profile has been computed
with 30 arcsec bins centered on the centroid of the image 
(RA=01 25 54; DEC=-01 21 05) out to 50 arcmin. The surface brightness profile
was fitted with a single $\beta$-model (Cavaliere \& Fusco-Femiano 1976)
plus a constant (to take into account the sky background
composed by Galactic foregrounds and residual CXB), with the 
software \textsc{PROFFIT} v1.3 (Eckert et al. 2011).
The best fitting model has a core radius $r_{\rm c} = 11.5\pm1.2$ arcmin (248.4 $\pm$ 26 kpc) 
and $\beta=0.67\pm0.06$ (1$\sigma$) for a $\chi^2$/dof=1.77.
In Fig.\,\ref{betamodel} we show the surface brightness profile of the X-ray 
emission of Abell~194, with the best fit $\beta$-model shown in blue.

For the determination of the spectral parameters representative of the core
properties we analyzed the \chandra\ ACIS-S observation of Abell~194 
(ObsID: 7823) with CIAO 4.7 and CALDB 4.6.8. All data were reprocessed from the 
\textsc{level=1} event file following the standard \chandra\ reduction threads
and flare cleaning. We used blank-sky observations to subtract the background
components and to account for variations in the normalization of the 
particle background we rescaled the blank-sky background template by the
ratio of the count rates in the 10-12 keV energy band. We extracted a spectrum
from a circular region of radius 1 arcmin around the centroid position.
We fitted the data with an \textsc{APEC} (Smith et al. 2001) model with 
ATOMDB code v2.0.2. in \textsc{XSPEC} v.12.8.2 (Arnaud 1996).
We fixed the Galactic column density at $N_{\rm{H}}=4.11 \times 10^{20}$ 
cm$^{-2}$ (Kalberla et al. 2005), the abundance is quoted in the solar
units of Anders \& Grevesse (1989) and we used the Cash statistic. 
The best fit model gives a temperature of $kT=2.4\pm0.3$ keV, an abundance of 
$0.27^{+0.15}_{-0.11}\,\rm{Z}_{\odot}$ and a \textsc{XSPEC} normalization of 
$1.0\pm0.1 \times 10^{-4}$ cm$^{-5}$ for a cstat/dof = 82/83.

Using the \rosat\ best fit $\beta$-model and the spectral parameters obtained
with \chandra\ the central electron density can be expressed by a simple
analytical formula (Eq. 2 of Ettori et al. 2004). In order to
assess the error we repeated the measurements after 1000 random realizations
of the normalization and the $\beta$-model parameters drawn from Gaussian 
distributions with mean and standard deviation set by the best-fit results.
We obtain a value for the central electron density of 
$n_{\rm 0}=(6.9\pm0.6) \times 10^{-4}$ cm$^{-3}$.
The distribution of the thermal electrons density with the distance from the cluster 
X-ray center $r$ was thus modeled with:
\begin{equation}
n_{\rm e}(r)=n_{\rm 0}\left(1+\frac{r^{\rm 2}}{r_{\rm c}^{\rm 2}}\right)^{\rm -\frac{3}{2}\beta}.
\label{beta}
\end{equation}

\section{Spectral aging analysis}
\label{spectrum}

\begin{table*}
\caption{Relevant parameters of the images smoothed to a resolution of 174\arcsec (2.9\arcmin), and 60\arcsec.}
\begin{center}
\begin{tabular} {lc|cccc|cccc} 
\hline
Frequency   & Orig. Beam        &  Conv. Beam    &  $\sigma_{\rm I}$ &  $\sigma_{\rm Q}$ &   $\sigma_{\rm U}$ & Conv. Beam    &  $\sigma_{\rm I}$ &  $\sigma_{\rm Q}$ &  $\sigma_{\rm U}$\\
(MHz)       &  (\arcsec$\times$\arcsec)        &  (\arcsec)     &              &    (mJy/beam)&               & (\arcsec)     &              &    (mJy/beam)&             \\
\hline
74 (VLSSr)   &  75$\times$75    & 174 &   260  &     $-$   & $-$    &         $-$            &       $-$       &    $-$ & $-$\\
330         &  21.8$\times$18.9 & 174 &   30         &     $-$   & $-$    &         $-$            &       $-$       &    $-$ & $-$\\
1443        &  59.3$\times$46.1 & 174 &   2.0        &    0.2  & 0.2  &     60     &      0.50     &  0.09& 0.07  \\
1465        &  52.8$\times$46.1 & 174 &   3.2        &    0.7  & 1.1  &     60     &      1.17     &  0.19& 0.28 \\
1515        &  49.8$\times$43.8 & 174 &   7.7        &    1.1  & 1.4  &     60     &      1.80     &  0.34& 0.35 \\
1630        &  52.8$\times$40.5 & 174 &   3.0        &    0.2  & 0.1  &     60     &      0.59     &  0.09& 0.08 \\
6600        &  174$\times$174   & 174 &   1.0        &    0.4  & 0.5  &           $-$          &       $-$   &  $-$ &  \\  
\hline
\multicolumn{10}{l}{\scriptsize Col. 1: Observing frequency; Col. 2: Original Beam; Col. 3: Smoothed Beam (174\arcsec); Col. 4, 5, 6: RMS noise of the I, Q, and U images smoothed to a resolution of 174\arcsec;}\\
\multicolumn{10}{l}{\scriptsize Col. 7: Smoothed Beam (60\arcsec); Col. 8, 9, 10: RMS noise of the I, Q, and U 
images smoothed to a resolution of 60\arcsec.}\\
\end{tabular}
\label{lowres}
\end{center}
\end{table*}

\begin{figure*}
\centering
\includegraphics[width=18 cm]{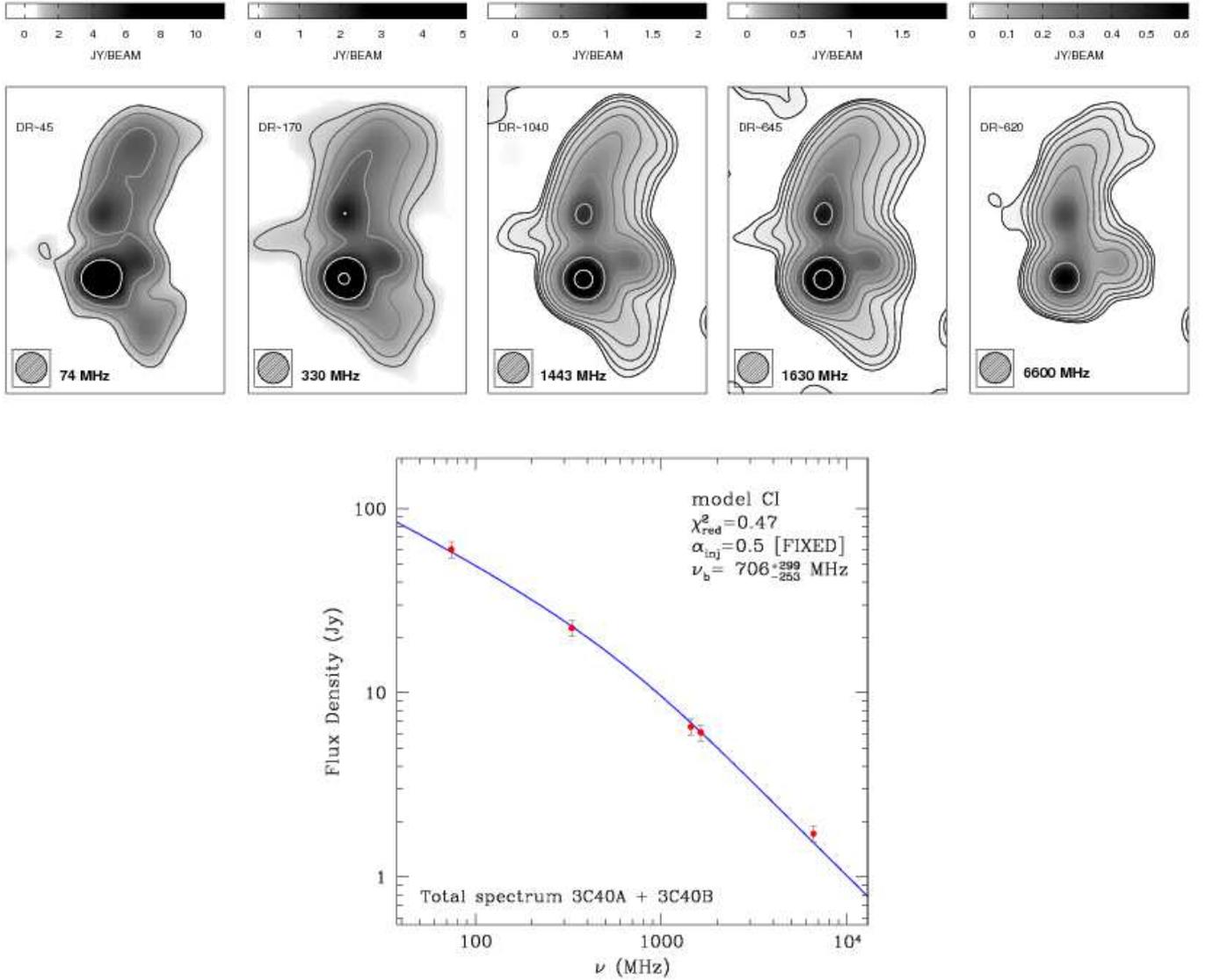}
\caption[]{
{\it Top}: Images of 3C\,40A and 3C\,40B at different frequencies, smoothed to a resolution of 2.9$'$.
The images are: VLSSr (74\,MHz; Lane et al. 2014), VLA P-Band (330 MHz),
VLA L-Band (1443 and 1630 MHz), and SRT (6600 MHz). The Dynamic Range (DR) of  
each image is shown on each of the top panels. 
{\it Bottom}: Total spectrum of sources 3C40\,A and 3C40\,B together. 
The blue line is the best fit of the CI model. 
}
\label{global}
\end{figure*}

\begin{figure*}
\centering
\includegraphics[width=18 cm]{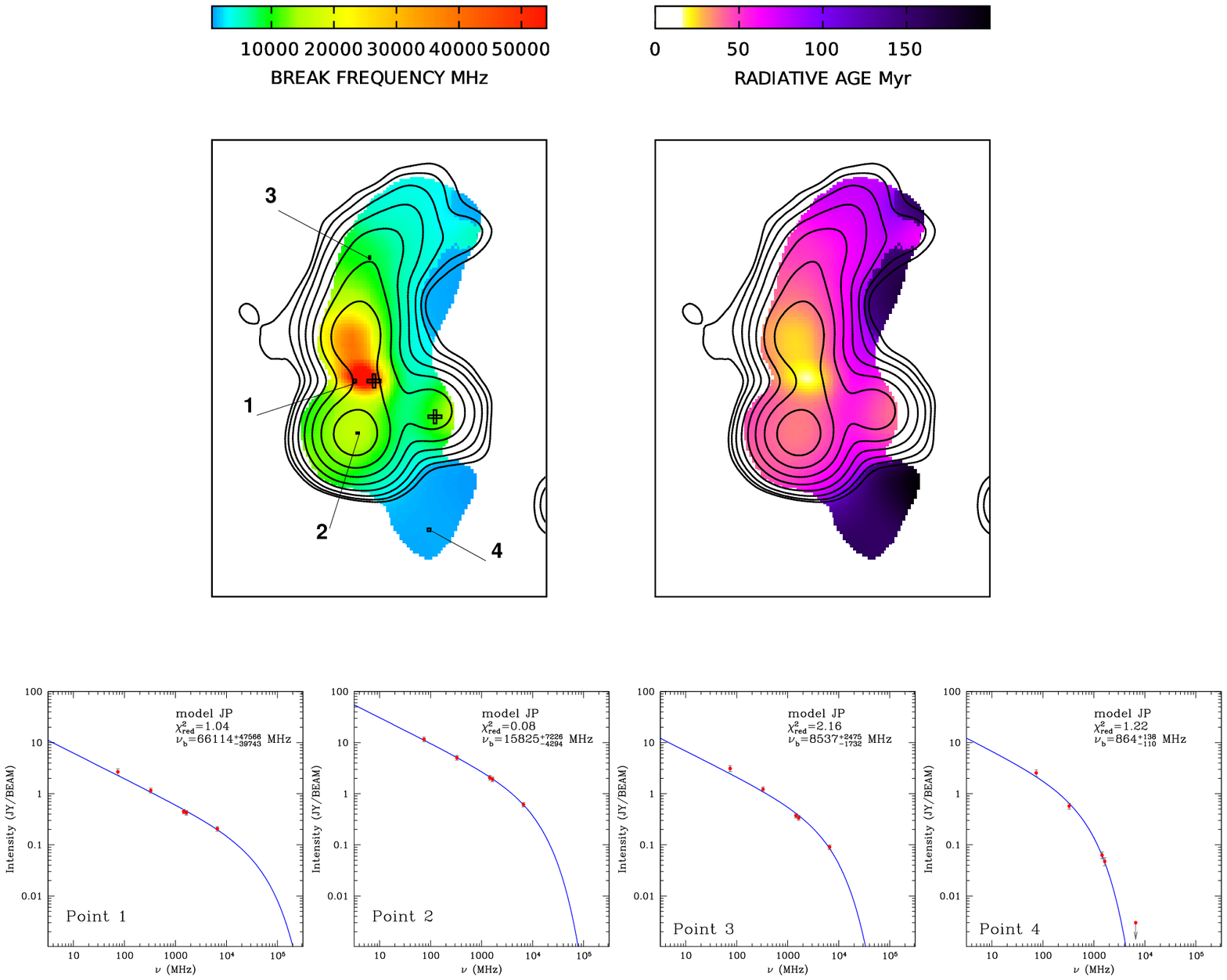}
\caption[]{
{\it Top-Left}: Map of break frequency, smoothed to a resolution of 2.9$'$ and derived with the pixel by pixel fit of the JP model 
in the regions where the brightness of the VLSSr image is above 5$\sigma_I$.
In the break frequency image there are about 15 independent beams.
Contours refer to the SRT image at 6600 MHz. Crosses indicate the optical 
core of 3C40\,A and 3C40\,B.
{\it Top-Right}: Map of the radiative age, smoothed to a resolution of 2.9$'$ and derived by applying in Eq.\,\ref{synage} the 
break frequency image and the minimum energy magnetic field. 
In the radiative age image there are about 15 independent beams.
Contours refer to the SRT image at 6600 MHz.
{\it Bottom}: Plots of the radio surface brightness as a function of the 
observing frequency $\nu$ at different locations.
The resulting fit of the JP model is shown as a blue line.
}
\label{mappe}
\end{figure*}

We investigated 3C\,40A and 3C\,40B at different 
frequencies in order to study in detail their spectral index behavior.

We analyzed the images obtained from the VLA Low-Frequency Sky Survey 
redux (VLSSr at 74\,MHz; Lane et al. 2014), 
the VLA archive data in P-Band (330\,MHz), the VLA archive data 
in L-Band (1443 and 1630 MHz), and the SRT (6600 MHz) data. 
We smoothed all the images to the same resolution as that of the SRT image 
(see top panels of Fig.\,\ref{global}). The relevant parameters of the
images smoothed to a resolution of 2.9\arcmin\, are reported in Table\,\ref{lowres}.

We calculated the flux densities of 3C40\,A and 3C40\,B together.
The total spectrum of the sources is shown in the 
bottom panel of Fig.\,\ref{global}. 
Although the radio sources in Abell~194 have a large angular extension,
the interferometric VLA data at 1443 and 1630 MHz seems do not suffer from missing flux problem.
To calculate the error associated to the flux densities, 
we added in quadrature the statistical noise and an additional uncertainty of 10\% 
to take into account for a different uv-coverage 
and a possible bias due to the different flux density scale of the data sets.

By using the software SYNAGE (Murgia et al. 1999), we fitted the integrated
spectrum with the continuous injection model (CI; Pacholczyk 1970).
The CI model is characterized by three free parameters: the injection spectral
index ($\alpha_{\rm inj}$), the break frequency ($\nu_{\rm b}$), and the flux normalization.
In the context of the CI model, it is assumed that
the spectral break is due to the energy losses of the relativistic electrons. 
For high-energy electrons, the energy losses
are primarily due to the synchrotron radiation itself and to
the inverse Compton scattering of the Cosmic Microwave
Background (CMB) photons.
During the active phase, the evolution of the integrated
spectrum is determined by the shift with time of $\nu_{\rm b}$ to lower
and lower frequencies. Indeed, the spectral break can be considered to be a clock 
indicating the time elapsed since the injection of the first electron population. 
Below and above $\nu_{\rm b}$, the spectral indices are respectively $\alpha_{\rm inj}$ and 
$\alpha_{\rm inj}$+0.5.

For 3C40\,A and 3C40\,B the best fit of the CI model to the observed radio spectrum yields 
a break frequency $\nu_{\rm b}\simeq700\pm280$ MHz.
To limit the number of free parameters, we fixed $\alpha_{\rm inj}=0.5$  which 
is the value of the spectral index calculated in the jets of 3C\,40B, at a 
relatively high resolution ($\simeq$ 20\arcsec), between the L-Band (1443 MHz) 
and the P-Band (330 MHz) images.

By using the images in the top panels of Fig.\,\ref{global}, we also studied the variation
pixel by pixel of the synchrotron spectrum along 3C\,40A and 3C\,40B.
3C\,40B is extended enough to investigate the spectral 
trend along its length.
A few plots showing the radio surface brightness as a function of the 
observing frequency at different locations of the 3C\,40B are shown in 
bottom panels of Fig.\,\ref{mappe}. The plots show some 
pixels, located in different parts of 3C\,40B, with representative spectral trend.

The morphology of the radio galaxy, and the trend of the spectral curvatures at different 
locations indicate that 3C\,40B is currently an active source whose synchrotron emission is dominated 
by the electron populations with GeV energies injected by the radio jets and accumulated during their 
entire lives (Murgia et al. 2011). 

We then fitted the observed spectra with the JP model (Jaffe \& Perola 1973),
which also has three free parameters $\alpha_{\rm inj}$, $\nu_{\rm b}$, and flux normalization 
like the CI model. The JP model however describes the
spectral shape of an isolated electron population with an isotropic distribution 
of the pitch angles injected at a specific instant in time with an initial 
power law energy spectrum with index $p=2\alpha_{\rm inj}+1$. 
According to the synchrotron theory (e.g. Blumenthal \& Gould 1970, Rybicki \& Lightman 1979), 
it is possible to relate the break frequency to the time elapsed since the start of 
the injection:

\begin{equation}
t_{\rm syn}= 1590 \frac{B^{0.5}}{(B^2+B_{\rm IC}^2) [(1+z)\nu_{\rm b}]^{0.5}}~\rm Myr,
\label{synage}
\end{equation}
 
where $B$ and $B_{\rm IC}=3.25(1+z)^2$ are the source magnetic field and the inverse Compton equivalent magnetic 
field associated with the CMB, respectively.
The resulting map of the break frequency derived by the fit of the JP 
model is shown in the top left panel of Fig.\,\ref{mappe}.
The break frequency is computed in the regions where the brightness of the VLSSr 
image smoothed to a resolution of 2.9 is above 5$\sigma_I$ ($>1.3$ Jy/beam).
The measured break frequency decreases systematically along the lobes of 3C\,40B  
in agreement with a scenario in which the oldest particles are 
those at larger distance from the AGN. 
The minimum break frequency measured in the faintest part of the lobes of 3C\,40B 
is $\nu_{\rm b}\simeq 850\pm 120$\,MHz, in agreement, within the errors, with the spectral 
break measured in the integrated spectrum of 3C40\,A and 3C40\,B. 

We derived the radiative age from Eq.\,\ref{synage}, using the minimum energy
magnetic field strength. The minimum energy magnetic field was
calculated assuming, for the electron energy spectrum, 
a power law with index $p=2\alpha_{\rm inj}+1=2$ and a 
low energy cut-off at a Lorentz factor $\gamma_{\rm low}=100$.
In addition, we assumed that non-radiating relativistic ions have the same 
energy density as the relativistic electrons. 
We used the luminosity at 330\,MHz, since radiative losses are less 
important at low frequencies. 
Modeling the lobes of 3C\,40B as two cylinders in the plane of the
sky, the resulting  minimum energy magnetic field, is $\simeq$1.8\,$\mu$G in the 
northern lobe and $\simeq$1.7\,$\mu$G in the southern lobe.
Assuming for the source's magnetic field $\langle B_{\rm min}\rangle=1.75\,\mu$G 
and the lowest measured break frequency of
$\nu_{\rm b}=850\pm 120$\,MHz, the radiative age of 3C\,40B is found to be
$t_{\rm syn}=157$$\pm$11\,Myrs. Sakelliou et al. (2008), applied a similar
approach and found a spectral age perfectly consistent with our result.
Furthermore, we note that the radiative age of 3C40\,B is consistent 
with that found in the literature for other sources of similar size and radio power
(see Fig. 6 of Parma et al. 1999).

In the top right panel of Fig.\,\ref{mappe}, we show the corresponding radiative age map 
of 3C\,40B obtained by applying in Eq.\,\ref{synage} the break frequency image and the minimum energy
magnetic field. 

In the computation above we assumed an electron to proton ratio $k=1$,
in agreement with previous works that we use for comparison.
We are aware that this ratio could be larger, and in particular
there could be local changes due to significant radiative losses
of electrons, with respect to protons, in oldest source regions.
We can estimate the impact of a different value of $k$ on the radiative age.
By keeping the assumptions adopted above and considering $k=100$,
the minimum energy magnetic field results a factor of three higher, and
the radiative age of 3C 40B is found to be $t_{\rm syn}=100\pm 7$\,Myrs,
which is slightly lower than the value derived by using $k=1$, however
the main results of the paper are not affected.

For extended sources like 3C40\,B, high-frequencies interferometric observations suffer the
so-called missing flux problem. On the other hand, the total intensity SRT image at 6600 MHz 
permitted us to investigate the curvature of the high-frequency spectrum
across the source 3C40\,B, which is essential to obtain a reliable aging analysis.
The above analysis is important not only to derive the radiative age of the radio galaxy but also
for the cluster magnetic field interpretation. 
Indeed, the break frequency image and the cluster X-ray emission model (see Sect.\,\ref{modeling}) are used 
in combination with the RM data to constrain the intra-cluster magnetic field power spectrum. 

\section{Faraday rotation analysis}
\label{rotmes}

In this section we investigate the polarization properties of the radio galaxies in Abell~194, by using data-sets at different resolutions.
First, we produced polarization images at 2.9$'$ resolution, useful to derive
the variation of the magnetic field strength in the cluster volume.
Second, we produced polarization images at higher resolution (19$''$ and 60$''$), useful to determine the correlation length of the magnetic field
fluctuations.

\subsection{Polarization data at 2.9\arcmin resolution}
\label{largescales}

\begin{figure*}
\centering
\includegraphics[width=15 cm]{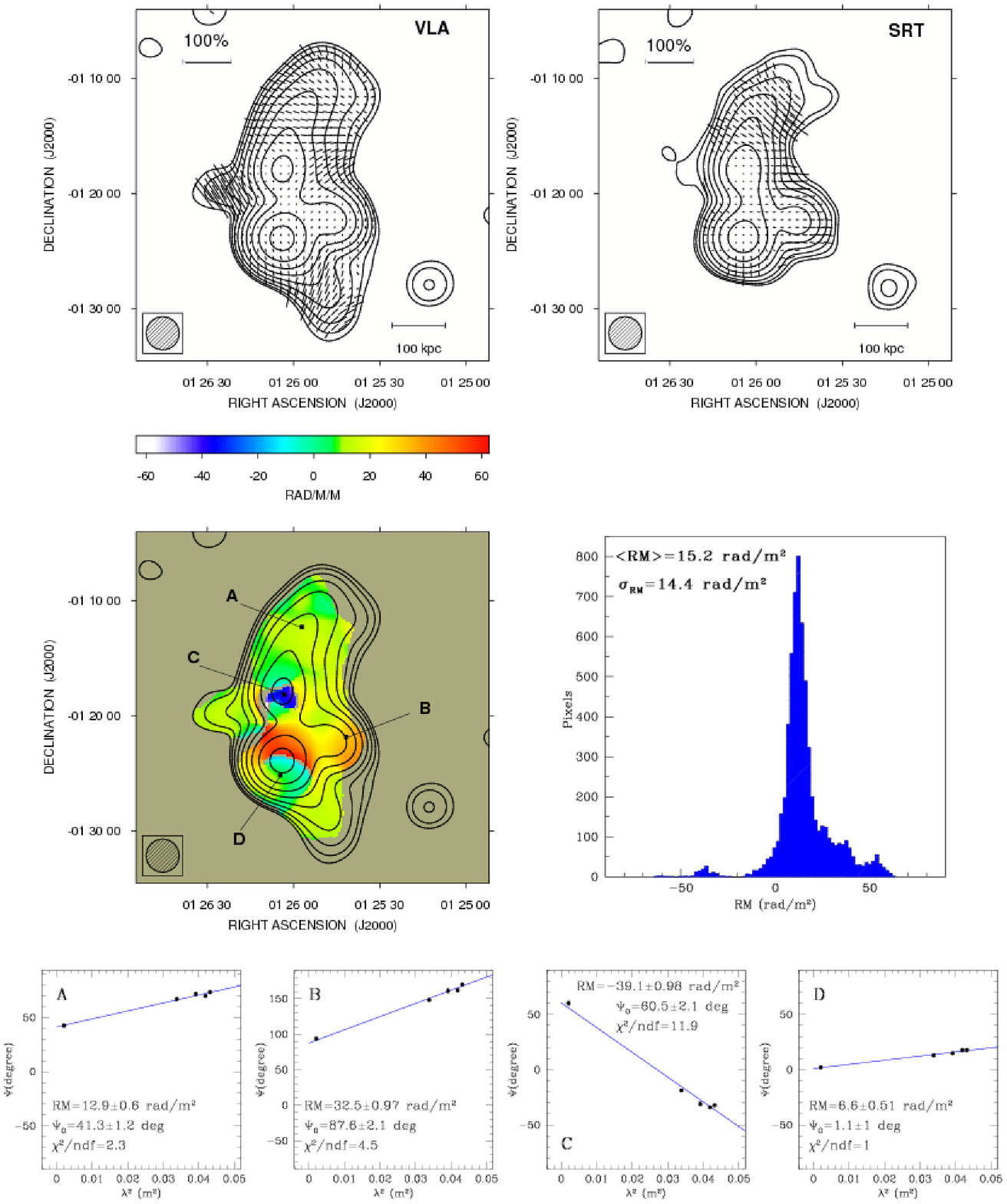}
\caption[]{
{\it Top}: Total intensity contours and polarization vectors of the radio sources 
in Abell~194. VLA image ({\it Left}) at 1443 MHz ($\sigma_{\rm I}$=2 mJy/beam),
smoothed with a FWHM Gaussian of 2.9\arcmin.
SRT image ({\it Right}) at 6600 MHz ($\sigma_{\rm I}$=1 mJy/beam), with a resolution of 2.9\arcmin.
The radio contours are drawn at 3$\sigma_{\rm I}$ and the rest are spaced by a factor of 2.
The lines give the orientation of the electric vector position angle (E-field)
and are proportional in length to the fractional polarization.
The vectors are traced only for those pixels where the total intensity
signal is above 5$\sigma_{\rm I}$, the error on the polarization angle is less than 10\deg, 
and the fractional polarization is above 3$\sigma_{\rm FPOL}$. 
{\it Middle-Left}: RM image calculated by using the images at 1443, 1465, 1515, 1630, and 6600 MHz, 
 smoothed with a FWHM Gaussian of 2.9\arcmin.
The color range is from $-$60 to 60 rad/m$^{2}$. Contours refer to the total 
intensity image at 1443 MHz as in the top-left panel.
{\it Middle-Right}: Histogram of the RM distribution.
{\it Bottom}: Plots of the position angle $\Psi$ as a function of $\lambda^2$ at 
four different source locations.
}
\label{rm_srt}
\end{figure*}

\begin{figure*}
\centering
\includegraphics[width=15.5 cm]{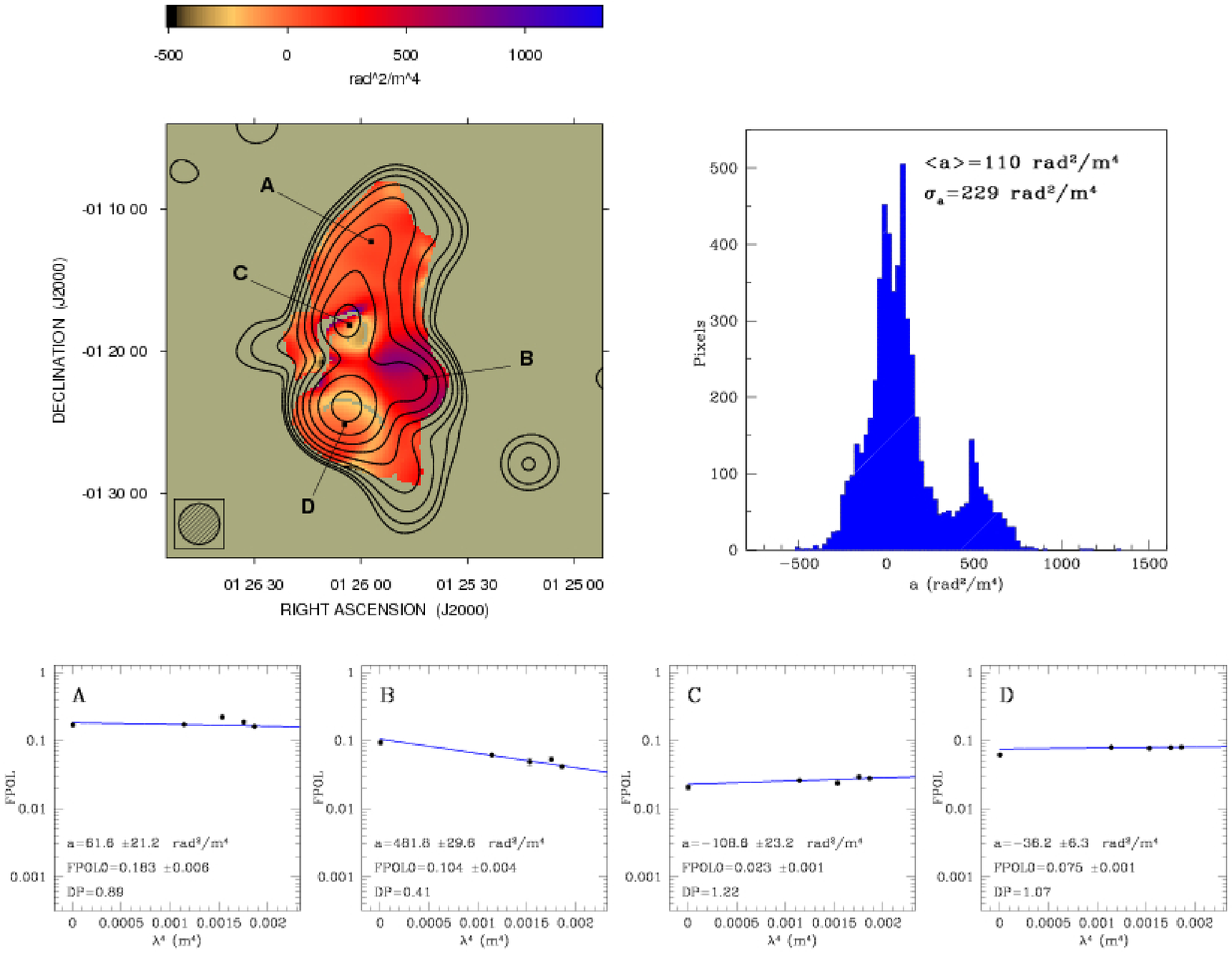}
\caption[]{
{\it Top-Left}: Burn-law $a$ in rad$^{2}$/m$^{4}$ calculated by using the images at 
1443, 1465, 1515, 1630, and 6600 MHz smoothed with a FWHM Gaussian of 2.9$'$. 
The image is derived from a fit of FPOL as a function of $\lambda^4$ 
following the Eq.\,\ref{burn}. The color range is from $-$500 to 1300 rad$^{2}$/m$^{4}$.
Contours refer to the total intensity image at 1443 MHz. 
The radio contours are drawn at 3$\sigma_{\rm I}$ and the rest are spaced by a factor of 2.
{\it Top-Right}: Histogram of the Burn-law $a$ distribution.
{\it Bottom}: Plots of fractional polarization FPOL a function of $\lambda^4$ at 
four different source locations.
}
\label{Fig_SRT_kburn}
\end{figure*}

In the top panels of Fig.\,\ref{rm_srt} we compare the polarization image 
observed with the SRT at 6600 MHz (right) with that
obtained with the VLA at 1443 MHz (left), smoothed with a FWHM 
Gaussian of 2.9$'$.

The fractional polarization at 1443 MHz is similar to that at 6600 MHz
being $\simeq 3-5$\% in 3C\,40A and in the brightest part of the radio 
lobes of 3C\,40B. The fractional polarization increases in the  
low surface brightness of the northern lobe of 3C\,40B where it is $\simeq$15\%.
The oldest low-brightness regions are visible both in total intensity and 
polarization at 1443 MHz but not at 6600 MHz, due to the sharp high-frequency cut-off
of the synchrotron spectrum. These structures, located east of 3C\,40B and in 
the southern lobe of 3C\,40B, show a fractional polarization
as high as $\simeq 20-50$\% at 1443 MHz (see Sect.\,\ref{modeling}).
 
By comparing the orientation of the polarization vectors at the different 
frequencies it is possible 
to note a rotation of the position angle.
We interpret this as due to the Faraday rotation effect.
The observed polarization angle $\Psi$ of a synchrotron 
radio source is modified from its intrinsic 
value $\Psi_{\rm 0}$ by the presence of a magneto-ionic Faraday screen
between the source of polarized emission and the observer.
In particular, the plane of polarization rotates according to:

\begin{equation}
\Psi=\Psi_{\rm 0}+{\rm RM}\times \lambda^2
\label{psi}
\end{equation}

\noindent
The RM is related to the plasma thermal 
electron density, $n_{\rm e}$, 
and magnetic field along the line-of-sight, $B_{\|}$ of 
the magneto-ionic medium, by the equation:
\begin{equation}
RM = 812\int\limits_0^L n_{\rm e} B_{\|} {\mathrm d}l ~~~{\rm rad/m^2}
\label{rm}
\end{equation}
where $B_{\|}$ is measured in $\mu$G, $n_{\rm e}$
in cm$^{-3}$, and $L$ is the depth of the screen in kpc. 

Following Eq.\,\ref{psi}, we obtained the RM image of Abell~194,  
by performing a fit pixel by pixel of the polarization angle images  
as a function of $\lambda^{2}$ using the FARADAY software (Murgia et al. 2004).
Given as input multi-frequencies images of Q and U, the software produces
the RM, the intrinsic polarization angle $\Psi_{\rm 0}$, and the corresponding 
error images. To improve the RM image,
the algorithm can be iterated in several self-calibration cycles. In the first cycle
only pixels with the highest signal-to-noise ratio are fitted. 
In the next cycles 
the algorithm uses the RM information in these high signal-to-noise pixels 
to solve the n$\pi$-ambiguity in adjacent pixels of lower signal-to-noise, in a 
similar method used in the PACERMAN algorithm by Dolag et al. (2005).

In the middle left panel of Fig.\,\ref{rm_srt}, we show the
resulting RM image at an angular resolution of 2.9\arcmin.
The values of RM range from $-$60 to 60 rad/m$^{2}$, with the higher absolute
values in the central region of the cluster in between the cores of
3C\,40B and 3C\,40A.
We obtained this image by performing the  $\lambda^{2}$ fit of the 
polarization angle images at 1443, 1465, 1515, 1630, and 6600 MHz.
The total intensity contours at 1443 MHz
are overlaid on the RM image. The RM image was derived
only in those pixels in which the following 
conditions were satisfied: the total intensity 
signal at 1443 MHz was above 3$\sigma_{\rm I}$, the resulting RM fit error was lower than 5 rad/m$^2$,
and in at least four frequencies the error in the polarization angle was lower than 10\deg. 

In the bottom panels of Fig.\,\ref{rm_srt}, we show a few plots of the position 
angle $\Psi$ as a function 
of $\lambda^2$ at different source locations. 
The plots show some 
pixels, located in different parts of the sources, with representative RM.
Thanks to the SRT image, we can effectively observe that
the data are generally quite well represented by a linear 
$\lambda^2$ relation over a broad $\lambda^2$ range, which 
supports the external Faraday screen hypothesis. 

We can characterize the RM distribution in terms of a mean 
$\langle RM \rangle$ and root mean square $\sigma_{\rm RM}$.
In the middle right panel of Fig.\,\ref{rm_srt}, we show
the histogram of the RM distribution. 
The mean and root mean square of the RM 
distribution are $\langle RM \rangle$=15.2 rad/m$^2$
and $\sigma_{\rm RM}$=14.4 rad/m$^2$, respectively. 
The mean fit error of the RM error image
is $\simeq$1.6 rad/m$^2$. The dispersion  
$\sigma_{\rm RM}$ of the RM distribution is higher than the mean RM error.
Therefore, the RM fluctuations observed in the image are 
significant and can give us information on the cluster magnetic 
field power spectrum.

For a partially resolved foreground, with Faraday rotation in a short-wavelength limit,
a depolarization of the signal due to the unresolved RM structures in the external screen
can be approximated, to first order, with the Burn-law (Burn 1966,
see also Laing et al. 2008 for a more recent derivation): 
\begin{equation}
\mathrm{FPOL=FPOL_0 \exp(-a\lambda^4)},
\label{burn}
\end{equation}

where $\mathrm{FPOL_0}$ is the fractional polarization at $\lambda=0$ and 
$a=2 \vert \nabla RM \vert^2 \sigma^2$ is related to the depolarization 
due to the 
RM gradient within the observing beam considered a circular Gaussian with 
FWHM=2$\sqrt{2ln2}\sigma$.

The effect of depolarization between two wavelengths $\lambda_1$ and $\lambda_2$ is usually
expressed in terms of the ratio of the degree of polarization 
DP=FPOL($\lambda_2$)/FPOL($\lambda_1$). The Burn-law can provide depolarization information 
by considering the data at all frequencies simultaneously. However, we note that the polarization
behaviour varies as the Burn-law at short wavelengths and goes to a simple
generic power-law form at long wavelengths (Tribble 1991).

Following Laing et al. (2008), in the top left panel of 
Fig.\,\ref{Fig_SRT_kburn} we show the image of the Burn-law exponent $a$ derived from 
a fit to the data. 
We obtain this image by performing the $\lambda^4$ fit of the
fractional polarization FPOL images at 1443, 1465, 1515, 1630, and 6600 MHz,
by using at least four of these frequencies in each pixel.
The values of $a$ range from $-$500 to 1300 rad$^{2}$/m$^{4}$.
The histogram of the Burn-law $a$ distribution is shown in the top 
right panel of Fig.\,\ref{Fig_SRT_kburn}. 
A large number of pixels have $a \simeq 0$ indicating no depolarization.
Significant depolarization ($a > 0$) is found close to 3C\,40A. Finally, a 
minority of pixels have $a < 0$. This may be due to the noise 
or the effect of Faraday rotation on a non-uniform distribution 
of intrinsic polarization (Laing et al. 2008).
We note that, indeed, re-polarisation of the signal is not unphysical 
and perfectly possible (e.g. Farnes et  al. 2014, Lamee et al 2016).
Example plots of $a$ as a function of $\lambda^4$ are 
shown in the bottom panels of Fig.\,\ref{Fig_SRT_kburn}.
The plots show the same pixels of Fig.\,\ref{rm_srt}. 
In these plots, the Burn-law 
gives adequate fits in the frequency range and resolution of these
observations.

We used the images presented in Fig.\,\ref{rm_srt}
and in Fig.\,\ref{Fig_SRT_kburn} to
investigate the variation of the magnetic field strength 
in the cluster volume. In addition, to measure the correlation length
of the magnetic field fluctuations, we used the archival VLA observations to 
improve the RM resolution in the brightest parts
of the sources.

\subsection{Polarization data at 19\arcsec\, and 60\arcsec\, resolution}

\begin{figure*}
\centering
\includegraphics[width=16 cm]{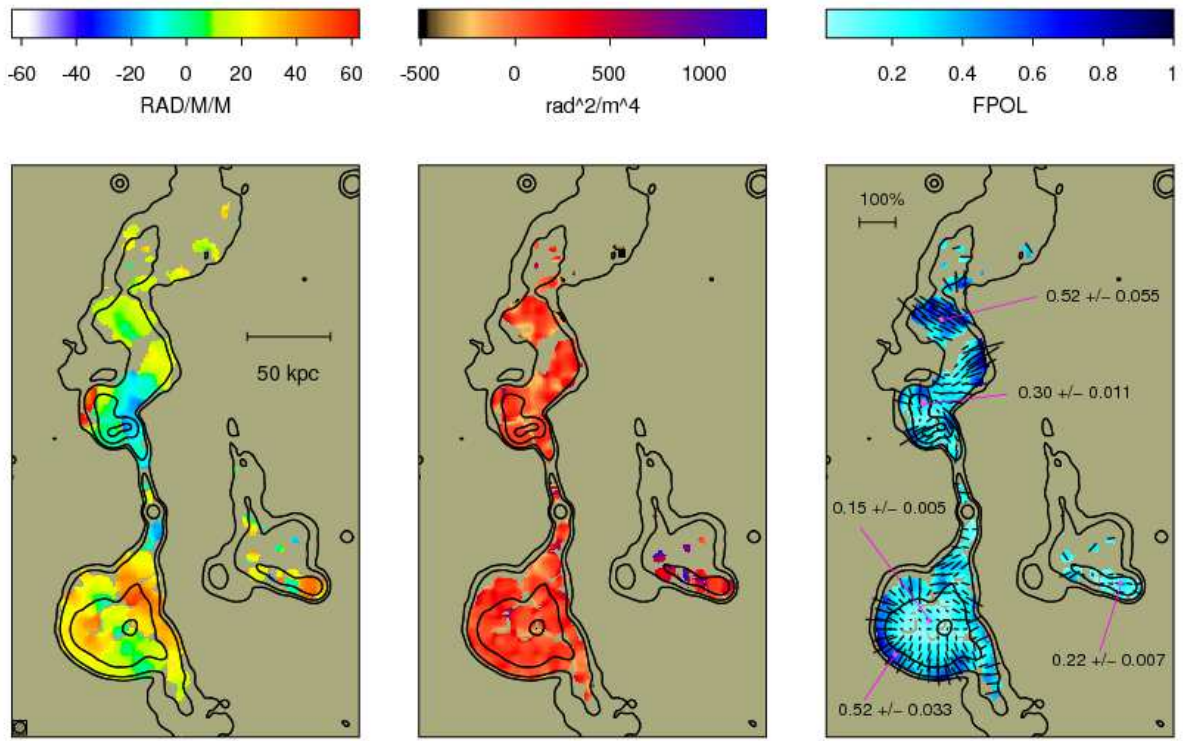}
\caption[]{
{\it Left}: Rotation measure image calculated by using the images 
at 1443, 1630, 4535, and 4885 MHz smoothed with a FWHM Gaussian of 19\arcsec. 
The color range is from $-$60 to 60 rad/m$^{2}$.  
Contours refer to the total intensity image at 1443 MHz. Levels are
1 (3$\sigma_{\rm I}$), 5, 30 and 100 mJy/beam.
{\it Middle}: 
Burn-law $a$ in rad$^{2}$/m$^{4}$ calculated by using the same images used to derive the RM.  
The image is derived from a fit of FPOL as a function of $\lambda^4$ 
following the Eq.\,\ref{burn}. The color range is from $-$500 to 1300 rad$^{2}$/m$^{4}$.
{\it Right}: Intrinsic polarization ($\Psi_{\rm 0}$ and FPOL$_{\rm 0}$), obtained by extrapolating to $\lambda=0$
both the RM (Eq.\,\ref{psi}) and the Burn-law (Eq.\,\ref{burn}).}
\label{Fig_MidRes}
\end{figure*}

In the left panel of Fig.\,\ref{Fig_MidRes} we show the resulting RM
image at an angular resolution of 19\arcsec. In the middle panel of 
Fig.\,\ref{Fig_MidRes}, we show the resulting Burn-law image, at the same
angular resolution. We derived these images by using VLA data 
at 1443, 1630, 4535, and 4885 MHz and by adopting the same 
strategy described in Sect.\,\ref{largescales} for the lower resolution images. 
The noise levels of the images at a resolution of 19\arcsec\, are given in Table\,\ref{midres}.
Qualitatively, the brightest parts of the 3C\,40A and 3C\,40B sources show RM and Burn-law images 
in agreement with what we find at lower resolution. However, 
at this resolution the two sources are well separated and the RM structures
can be investigated in a finer detail. In particular, the RM distribution seen over the two
radio galaxies is patchy, indicating a cluster magnetic field
with turbulent structures on scales of a few kpc. 

\begin{table*}
\caption{Relevant parameters of the total intensity and polarization images at a resolution of 19\arcsec.}
\begin{center}
\begin{tabular} {lccccc} 
\hline
Frequency & Orig. Beam      &   Conv. Beam         &  $\sigma_{\rm I}$ &  $\sigma_{\rm Q}$ & $\sigma_{\rm U}$   \\
(MHz)     &  (\arcsec$\times$\arcsec)  &  (\arcsec)     &  (mJy/beam)  &    (mJy/beam)    &  (mJy/beam) \\
\hline
 1443     &  16.8$\times$15.2 & 19   &   0.34       &   0.12   & 0.13 \\
 1630     &  15.0$\times$13.7 & 19   &   0.26       &   0.08   & 0.11 \\
 4535     &  19.7$\times$15.3 & 19   &   1.50       &   0.11   & 0.11  \\
 4885     &  18.1$\times$14.0 & 19   &   1.33       &   0.17   & 0.09  \\
\hline
\multicolumn{6}{l}{\scriptsize Col. 1: Observing frequency; Col. 2: Original Beam; Col. 3: Smoothed Beam;}\\ 
\multicolumn{6}{l}{\scriptsize Col. 4, 5, 6: RMS noise of the $I$, $Q$, and $U$ images at 19$''$ of resolution.}\\
\multicolumn{6}{l}{\scriptsize We note that the noise levels of the images at 4535 and 4885 MHz contain}\\
\multicolumn{6}{l}{\scriptsize the PBcore correction, being obtained with the task FLATN of three}\\
\multicolumn{5}{l}{\scriptsize different pointings.}\\
\end{tabular}
\label{midres}
\end{center}
\end{table*}

\begin{figure*}
\centering
\includegraphics[width=16 cm]{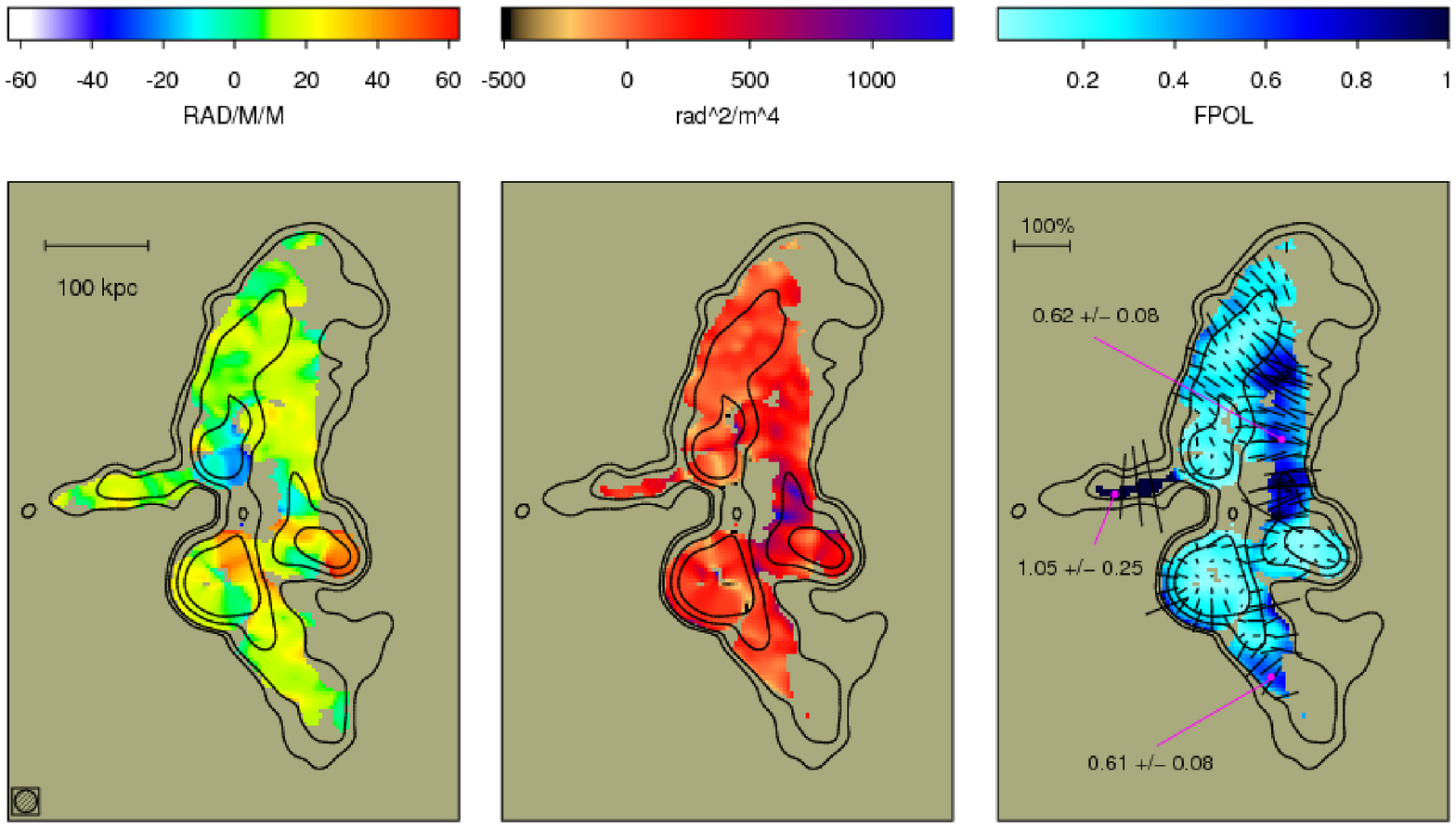}
\caption[]{
{\it Left}: Rotation measure image calculated by using the images 
at 1443, 1465, 1515, and 1630 MHz smoothed with a FWHM Gaussian of 60$''$.
The color range is from $-$60 to 60 rad/m$^{2}$.
Contours refer to the total intensity image at 1443 MHz. Levels are
1.5 (3$\sigma_{\rm I}$), 5, 30 and 100 mJy/beam.
{\it Middle}: 
Burn-law $a$ in rad$^{2}$/m$^{4}$ calculated by using the same images used to derive the RM.  
The image is derived from a fit of FPOL as a function of $\lambda^4$ 
following the Eq.\,\ref{burn}. The color range is from $-$500 to 1300 rad$^{2}$/m$^{4}$.
{\it Right}: Intrinsic polarization ($\Psi_{\rm 0}$ and FPOL$_{\rm 0}$), obtained by extrapolating to $\lambda=0$
both the RM (Eq.\,\ref{psi}) and the Burn-law (Eq.\,\ref{burn}).}
\label{Fig_LowRes}
\end{figure*}

In the left and middle panel of Fig.\,\ref{Fig_LowRes} we show the resulting RM
and Burn-law images at an angular resolution of 60\arcsec. We derived these images by using VLA data 
at 1443, 1465, 1515, and 1630 MHz and by adopting the same strategy described in Sect.\,\ref{largescales}. 
The noise level of the images at a resolution of 60\arcsec\, are given in Table\,\ref{lowres}.
The RM image at 60\arcsec\, is consistent with that at 2.9\arcmin, although obtained
in a narrower $\lambda^2$ range.

We note that low polarization structures may potentially causing
images with some artefacts. However, these effects are taken into account 
in the magnetic field modeling since simulations and data 
are filtered in the same way. The data at 19\arcsec\, and 60\arcsec\, are used to derive realistic models
for the intrinsic properties of the radio galaxies. These models are used 
to produce synthetic polarization images at arbitrary frequencies 
and resolutions
in our magnetic field modeling (see Sect.\,\ref{modeling}). 

In the right panel of Fig.\,\ref{Fig_MidRes} we show the intrinsic polarization  
image at 19\arcsec\, of 3C\,40A and 3C\,40B. The intrinsic 
polarization ($\Psi_{\rm 0}$ and FPOL$_{\rm 0}$),
was obtained by extrapolating to $\lambda=0$ both the RM (Eq.\,\ref{psi}) and the Burn-law (Eq.\,\ref{burn}).
The image reveals regions of high fractional polarization. The panel shows 
some representative values of fractional polarization with the corresponding 
uncertainty. We note that the southern lobe of 3C40B shows to the 
south-east an outer rim of high fractional polarization ($\simeq 50\%$).
These polarisation rims are commonly observed at the boundary of the lobes
of both low (e.g. Capetti et al. 1993) and high (e.g. Perley \& Carilli 1996) 
luminosity radio galaxies. The rim-like morphology of high fractional polarisation is an expected feature usually interpreted as the compression of the lobes magnetic field along the contact discontinuity. The normal field components near the edge are suppressed and only tangential components survive giving rise to very high fractional polarisation levels.

This image provides a description of the intrinsic polarization properties 
in the brightest part of the sources.
However, for modeling at large distances from the cluster center we need information also in 
the fainter parts of the sources.
In this case we obtained the information from the data at a resolution of 60$''$.
In the right panel of Fig.\,\ref{Fig_LowRes} we show the intrinsic polarization  
image at 60$''$, derived by extrapolating to $\lambda=0$ both the RM formula 
and the Burn-law. The two images are in good agreement in the common parts. 
In addition, they complement the intrinsic polarization information in different 
sampling regions of the sources.
The intrinsic ($\lambda=0$) fractional polarization images shown in 
Fig.\,\ref{Fig_MidRes} and Fig.\,\ref{Fig_LowRes}, confirm 
very low polarization levels in coincidence with the peak intensity in the southern lobe, 
in agreement with the small length measured for the fractional polarization vectors
overimposed to the polarization image obtained with the SRT at 6600 MHz and shown in Fig.\,\ref{Fig_SRT_Pol}.
Furthermore, the higher S/N ratio of the low resolution image, reveals a few 
steep spectrum diffuse features not observable in Fig.\,\ref{Fig_MidRes}.
In particular: a trail to the east (see Sakelliou et al. 2008, for an interpretation), the northern part
of the tail of 3C40A, that overlaps in projection to the tip of the northern lobe of 3C40B, and a tail extending from 
the southern lobe of 3C40B. A common property of these features is that they are all highly polarised as indicated in the figure.
These are likely very old and relaxed regions characterized by an ordered magnetic field. The combination of these two effects
could explain the high fractional polarization observed.

\subsection{Polarization results summary}

The SRT high-frequency observations are essential to
determine the large scale structure of the RM in Abell~194. 
While the angular size of the lobes shown in Fig.\,\ref{Fig_MidRes} are comparable with the largest angular scale 
detectable by the VLA in D configuration at C-Band ($\simeq$5\arcmin),
the full angular extent of 3C40\,B is so large ($\simeq$20\arcmin) that the intrinsic structure of the U and Q
parameters can be properly recovered only with single-dish observations. Indeed, the SRT observation at
6600 MHz is fundamental because it makes possible to ensure that the polarization position angles follow
the $\lambda^2$ law over a broad range of wavelengths thus confirming the hypothesis that the RM originates
in an external Faraday screen.

The observed RM fluctuations in Abell~194 appear  
similar to those observed for other sources in comparable environments, e.g.
NGC\,6251, 3C\,449, NGC\,383, (Perley et al. 1984, Feretti et al. 1999, Laing et al. 2008), but are 
much smaller in amplitude than in rich galaxy clusters (e.g. Govoni et al. 2010).

We summarize the RM results obtained at the different resolutions 
in Table\,\ref{rmtab}.
The values of $\langle{\mathrm RM}\rangle$ given in Table\,\ref{rmtab} include
the contribution of the Galaxy. 
We determined the RM Galactic contribution by using the results 
by Oppermann et al. (2015), mostly based on the catalog by 
Taylor et al. (2009), which contains the RM for 34 radio sources
within a radius of 3 deg from the cluster center of Abell~194.
Within this radius, the RM Galactic contribution results 8.7$\pm$4.5 rad/m$^2$. 
Therefore, half of the $\langle RM \rangle$ value measured 
in the Abell~194 sources is Galactic in origin.

In 3C\,40B the two lobes show a similar RM and Burn-law $a$, this is in contrast to what was found 
in 3C\,31 (Laing et al. 2008), where asymmetry between the north and 
south sides is evident. While 3C\,31 is inclined with respect the plane of the
sky and the receding lobe is more rotated than the approaching lobe (Laing-Garrinton 
effect; Laing 1988, Garrinton et al. 1988), in Abell~194, our results seem
to indicate that the two lobes of 3C\,40B lay on the plane of the 
sky, and thus they are affected by a similar physical depth of the Faraday
screen. It is also interesting to note that 3C\,40A is located at a projected distance
from the cluster center very similar to that of the southern lobe of 3C\,40B, but 3C\,40A
is more depolarized than the southern lobe of 3C\,40B.
This behavior may be explained if 3C\,40A is located at a deeper
position along the line of sight than 3C\,40B. Thus, passing across 
a longer path, the signal is subject 
to an higher depolarization, due to the higher RM. 
Other explanations may be connected to the presence of the X-ray cavity in the 
southern lobe of 3C\,40B which might reduce its depolarization with 
respect to 3C\,40A.

The FPOL trend with $\lambda^4$ and 
the polarization angle trend with $\lambda^2$  would argue in favor of 
an RM produced by a foreground Faraday screen 
(Burn 1966, Laing 1984, Laing et al. 2008). 
In addition, the presence of a large X-ray cavity formed by 
the southern radio lobe arising from 3C\,40B (Bogd{\'a}n et al. 2011), 
supports the scenario in which the rotation of the polarization plane is likely to occur 
entirely in the intra-cluster medium, since the radio lobe is likely voided of
thermal gas.

In a few cases (Guidetti et al. 2011) the RM morphologies 
tend to display peculiar anisotropic RM structures. 
These anisotropies appear highly ordered (‘RM bands’) with iso-contours 
orthogonal
to the axes of the radio lobes and are likely related to the interaction 
between the sources and their surroundings. 
The RM images of Abell~194 show patchy patterns without any obvious
preferred direction, in agreement with many of the published RM images.
Therefore, in Abell~194, the standard picture that the Faraday effect is 
due to the turbulent intra-cluster medium and not affected 
by the presence of the radio source itself, 
is self-consistent.

These above considerations suggest that the effect of the external
Faraday screen is dominant over the internal Faraday rotation, if any.

As shown in optical and X-ray analyses
(see Sect.\,\ref{optical} and\,\ref{xray}), Abell~194 is not undergoing a major merger event. Rather
it is likely accreting several smaller clumps. Under these conditions, we do not expect injection of turbulence on cluster-wide scales in the intra-cluster medium.
However, we may expect injection of turbulence on scales of a few hundred
kpc due to the accretion flows of small groups of galaxies, 
the motion of galaxy cluster members, and the expansion of radio lobes.
This turbulent energy will cascade on smaller scales and will be dissipated
finally heating the intra-cluster medium. The footprint of this cascade
could be observable on the intra-cluster magnetic field structure through
the RM images. In this case, the patchy structure characterizing the RM images 
in Figs.\,\ref{rm_srt},\,\ref{Fig_MidRes}, and \,\ref{Fig_LowRes}, can be interpreted as a signature of the 
turbulent intra-cluster magnetic field and the values of the $\sigma_{\rm RM}$
and $\langle RM \rangle$ can be used to constrain the strength and the 
structure of the magnetic field.

\begin{table*}
\caption{Rotation measure results.}
\centering
\begin{tabular} {c r r r c r r} 
\hline
Source & Resolution  & Distance & N  & $\langle \mathrm{Errfit_{\rm RM}} \rangle $   & $\langle RM \rangle$   &  $\sigma_{\rm RM}$   \\
       &  (\arcsec)   &  (kpc)   &    & (rad/m$^{2}$)   & (rad/m$^{2}$)          &  (rad/m$^{2}$)      \\
\hline  
3C\,40B/3C\,40A  &   174  & -        & 19   & 1.6            & 15.2     & 14.4                \\ \hline
3C\,40B/3C\,40A  &   60   & -        & 72   & 2.4            & 14.8     & 12.4                \\ \hline 
3C\,40B/3C\,40A  &   19   & -        & 146  & 1.4            & 16.5     & 15.8        \\
3C\,40B        &   19   & 35       & 135  & 1.4            & 15.9     & 15.6        \\
3C\,40A        &   19   & 70      & 11   & 1.9            & 22.7     & 17.9         \\          \hline
\multicolumn{7}{l}{\scriptsize Col. 1: Source; Col. 2 Resolution; Col. 3: Projected distance of the
source core from the X-ray centroid;}\\
\multicolumn{7}{l}{\scriptsize Col. 4: Number of beams over which the RM has been computed;}\\  
\multicolumn{7}{l}{\scriptsize Col. 5: Mean value of the RM fit error;}\\ 
\multicolumn{7}{l}{\scriptsize Col. 6: Mean of the RM distribution; Col. 7: RMS of the RM distribution.}\\ 
\end{tabular}
\label{rmtab}
\end{table*}

\section{Intra-cluster magnetic field characterization}
\label{modeling}

We constrained the magnetic field power spectrum in Abell~194
by using the polarization information (RM and fractional polarization)
derived for the radio galaxies 3C\,40A and 3C\,40B.
A realistic description of the cluster magnetic fields must provide
simultaneously a reasonable representation of the following observables:\\

$-$ the RM structure function defined by:
\begin{equation}
S(\Delta x, \Delta y)=\langle [RM(x,y)-RM(x+\Delta x,y+\Delta y)]^2\rangle_{(x,y)},
\end{equation}
where $=\langle \rangle_{(x,y)}$ indicates that the average is taken
over all the positions $(x,y)$ in the RM image. The structure function $S(\Delta r)$ was
then computed by azimuthally averaging $S(\Delta x,\Delta y)$ over annuli of
increasing separation $\Delta r=\sqrt{\Delta x^2+\Delta y^2}$;\\

$-$ the Burn-law (see Eq.\,\ref{burn});\\

$-$ the trends of $\sigma_{RM}$ and $\langle RM \rangle$ against the distance from the cluster center.\\

The RM structure function and the Burn-law are better investigated by using the data at higher resolution (19$''$),
while the trends of $\sigma_{RM}$ and $\langle RM \rangle$ with the distance from the cluster 
center are better described by the data at 
lower resolution (2.9$'$).

Our modeling is based on the assumption that the Faraday rotation 
occurs entirely in the intra-cluster medium. In particular, we 
supposed that there is no internal Faraday rotation inside the radio
lobes. 

To interpret the RM and the fractional polarization data, 
we need a model for the spatial distribution of the thermal electron density
and of the intra-cluster magnetic field.

For the thermal electrons density profile we assumed the 
$\beta$-model derived in Sect.\,\ref{xray}, 
with $r_{\rm c}$ = 248.4 kpc, $\beta$=0.67, 
and $n_{\rm 0}$=6.9 $\times$ 10$^{-4}$ cm$^{-3}$.

For the power spectrum of the intra-cluster magnetic field fluctuations
we adopted a power-law with index $n$ of the form

\begin{equation}
\vert B_{\rm k}\vert ^{\rm 2}\propto k^{\rm -n}
\label{eq3}
\end{equation}

\noindent
in the wave number range from $k_{\rm min}=2\pi/\Lambda_{\rm max}$ to 
$k_{\rm max}=2\pi/\Lambda_{\rm min}$. 

Moreover, we supposed that the power-spectrum normalization varies with the 
distance from the cluster center such that the average magnetic field strength at a given radius 
scales as a function of the thermal gas density according to
\begin{equation}
\langle  B(r)\rangle=\langle  B_{\rm 0}\rangle\left[\frac{n_{\rm e}(r)}{n_{\rm 0}}\right]^{\rm \eta},
\label{eta}
\end{equation}
where $\langle B_{\rm 0}\rangle$ is the average magnetic field strength at the center 
of the cluster\footnote{Since the simulated magnetic field components follow a Gaussian distribution, the total magnetic field is distributed according to a Maxwellian.
In this work we quote the average magnetic field strength which is related to the rms magnetic field strength through: $\langle B\rangle=\sqrt{8/(3\pi-8)} B_{\rm rms}$}, and $n_{\rm e}(r)$ is the thermal electron gas density radial profile. 
The magnetic field fluctuations are considered isotropic on scales $\Lambda \gg \Lambda_{\rm max}$,
i.e. the fluctuations phases are completely random in the Fourier space.

Overall, our magnetic field model depends on the five parameters
listed in Table\,\ref{parameters}:\\ 
\noindent
$(i)$ the strength at the cluster center $\langle B_{\rm 0}\rangle$;\\
$(ii)$ the radial index $\eta$;\\
$(iii)$ the power spectrum index $n$;\\
$(iv)$ the minimum scales of fluctuation $\Lambda_{\rm min}$;\\
$(v)$ the maximum scales of fluctuation $\Lambda_{\rm max}$.\\
As pointed out in Murgia et al. (2004), there are a number of degeneracies between these parameters. 
In particular, the most relevant to us are the degeneracy between $\langle B_0\rangle$ and $\eta$, 
and between $\Lambda _{\rm min}$, $\Lambda _{\rm max}$, and $n$.

Fitting all of these five parameters simultaneously would be the best way to proceed but, due to 
computational burden, this is not feasible. The aim of this work is to constrain the magnetic field
radial profile, therefore we proceeded in two steps. First, we performed two-dimensional simulations
to derive the magnetic field shape of the power spectrum. Second, we performed 
three-dimensional simulations varying the values of $\langle B_{\rm 0}\rangle$ and $\eta$ and
derive the magnetic field profile that best reproduce the RM observations.

We focused our analysis on $\langle B_0\rangle$, $\eta$, and $\Lambda _{\rm max}$.
In our modeling, $\Lambda _{\rm max}$ should represents
the injection scale of turbulent energy. Since Abell~194 is a rather relaxed cluster, we expect a $\Lambda _{\rm max}$
to be a hundred of kpc or less. The turbulent energy will cascade down to smaller and smaller scales
until it is dissipated. Determining the slope of the magnetic field power spectrum and the outer scale
of the magnetic field fluctuations is not trivial because of the degeneracy between these parameters (see e.g.
Murgia et al. 2004, Bonafede et al. 2010). To reduce the number of free model parameters, 
on the basis of the Kolmogorov theory for a turbulent medium, we fixed the slope of the power-law 
power spectrum to $n=11/3$.
In addition, we fixed $\Lambda_{\rm min} = 1$ kpc. We note that a different $\Lambda_{\rm min}$ has a negligible impact on 
these simulation results since, for a Kolmogorov spectral index, most of the magnetic field power is on 
larger scales and our observations do not resolve sub-kpc structures. 

\begin{table*}
\caption{Parameters of the magnetic field model.}    
\centering          
\begin{tabular}{c l l c} 
\hline
    Parameter                & Description                   & Value            & Means of investigation \\ 
\hline                        
  $n$                        & Power spectrum index, $\vert B_{\rm k}\vert ^{\rm 2} \propto k^{\rm -n}$& Fixed; $n=11/3$ & $-$\\
  $\Lambda_{\rm min}$          & Minimum scale of fluctuation, $\Lambda_{\rm min}=2\pi/k_{\rm max}$& Fixed; $\Lambda_{\rm min}=1$ kpc & $-$\\
  $\Lambda_{\rm max}$          & Maximum scale of fluctuation, $\Lambda_{\rm max}=2\pi/k_{\rm min}$& Free     & 2-D simulations\\
  $\langle  B_{\rm 0}\rangle$  & Strength at the cluster center & Free            & 3-D simulations\\
  $\eta$                     & Radial index, $\langle  B(r)\rangle=\langle  B_{\rm 0}\rangle\left(\frac{n_{\rm e}(r)}{n_{\rm 0}}\right)^{\rm \eta}$ & Free & 3-D simulations\\
\hline  
\multicolumn{4}{l}{\scriptsize Col. 1: Parameters of the magnetic field model; Col. 2: Description of the parameters;}\\ 
\multicolumn{4}{l}{\scriptsize Col. 3: Value of the parameters in the simulations (Free or Fixed);}\\
\multicolumn{4}{l}{\scriptsize Col. 4: Type of simulation used to constrains the parameters (two-dimensional, three-dimensional)}\\ 
\end{tabular}
\label{parameters} 
\end{table*}

In particular, by using the 19$''$ resolution data set, we performed a two-dimensional analysis to constrain
the maximum scale of the magnetic field fluctuations $\Lambda_{\rm max}$ and by using the 2.9$'$ resolution 
data set, we performed three-dimensional numerical simulations to constrain the strength of the magnetic 
field and its scaling with the gas density.
In both cases, we used the FARADAY code (Murgia et al. 2004), to produce 
simulated RM and fractional polarization images and to compare them with the observed
ones. The simulated images were produced by integrating numerically
the gas density and the magnetic field product along the line-of-sight.
We considered the sources 3C40\,A and 3C40\,B lying in the plane of the sky 
(not inclined) and the limits of the integrals were [0,1 Mpc], i.e.
with the sources located at the cluster center.
In the simulated RM images we took into account the Galactic RM contribution,  
which is supposed to be uniform over the radio sources 3C40\,A and 3C40\,B.

Due to the heavy window function imposed by the limited projected size
of the radio galaxies with respect to the size of the cluster,
for a proper comparison with observations,
the simulated fractional polarization and RM images 
are filtered like the observations before comparing them to data.
The two-dimensional and three-dimensional simulations are filtered following these steps:\\

$(1)$ We created a source model at 19$''$ (for the two-dimensional analysis)
and at 60$''$ (for the three-dimensional analysis), by considering the CLEAN components (CC model)
of the images at 1443 MHz. For each observing frequency, we then derived  
the expected total intensity images I($\nu$) obtained by associating at each CLEAN component
the corresponding flux density calculated 
by assuming a JP model with $\alpha_{\rm inj}=0.5$ and $\nu_{\rm b}$ taken from the observed break frequency 
image (see Fig.\,\ref{mappe} and Sect.\,\ref{spectrum});\\

$(2)$ We used the images in the right panels of Figs.\,\ref{Fig_MidRes} and \,\ref{Fig_LowRes},  
obtained by extrapolating the Burn-law at $\lambda$=0, as a model of intrinsic fractional 
polarization FPOL$_{\rm 0}$,
thus removing the wavelength-dependent depolarization due to the foreground Faraday screen.
In this way, FPOL$_{\rm 0}$ accounts for the intrinsic disorder of the magnetic field inside the radio source,
which may vary in general from point to point.  
For each observing frequency, we then derived the expected ``full-resolution''
fractional polarization FPOL($\nu$) obtained by associating at each CLEAN component the intrinsic 
fractional polarization corrected for the aging effect derived from the break frequency image
shown in Fig.\,\ref{mappe}: 
\begin{equation}
\mathrm{FPOL(\nu)=FPOL_{0} \times FPOL_{theor}(\nu/\nu_b)}.
\end{equation}
Murgia et al. (2016) pointed out that, as the synchrotron plasma ages, the theoretical
fractional polarization\footnote{The theoretical fractional polarization level $FPOL_{\rm theor}(\nu/\nu_{\rm b})$, expected for  $\alpha_{\rm inj}=0.5$, 
is shown in Fig. 15 of Murgia et al. (2016).} approaches 100\% for $\nu \gg \nu_{\rm b}$. The canonical
value $FPOL_{\rm theor}=(3 p +3)/(3 p + 7)$, valid for a power-law emission 
spectrum, is obtained in the limit $\nu \ll \nu_{\rm b}$.\\

$(3)$ For the intrinsic polarization angle $\Psi_{\rm 0}$, we assigned at each CLEAN component 
the values taken from the images in the right panels of Figs.\,\ref{Fig_MidRes} and \,\ref{Fig_LowRes}, 
obtained by extrapolating the $\Psi$ at $\lambda$=0;\\

$(4)$ By using I($\nu$), FPOL($\nu$), and $\Psi_{\rm 0}$, we produced the expected Q($\nu$) and U($\nu$) images corresponding 
to the simulated RM, obtained by employing Eq.\,\ref{psi};\\

$(5)$ We added a Gaussian noise such that of the observations to the I($\nu$), Q($\nu$), and U($\nu$) images
and we smoothed them with the FWHM beam of the observations  
used in the two-dimensional and three-dimensional analysis  (19$''$ or 2.9$'$). 
In this way we have taken into account the beam depolarization of the polarized signal;\\

$(6)$ Finally, we derived the fractional polarization images
FPOL($\nu$)=P($\nu$)/I($\nu$) from synthetic I($\nu$), Q($\nu$), and U($\nu$) images and we produced
the synthetic RM image by fitting pixel by pixel the polarization angle images $\Psi$($\nu$) 
versus the squared wavelength using exactly the same algorithm and strategy described 
in Sect.\,\ref{rotmes} as if they were actual observations.\\

\begin{figure*}[!]
   \centering
  \includegraphics[width=17 cm]{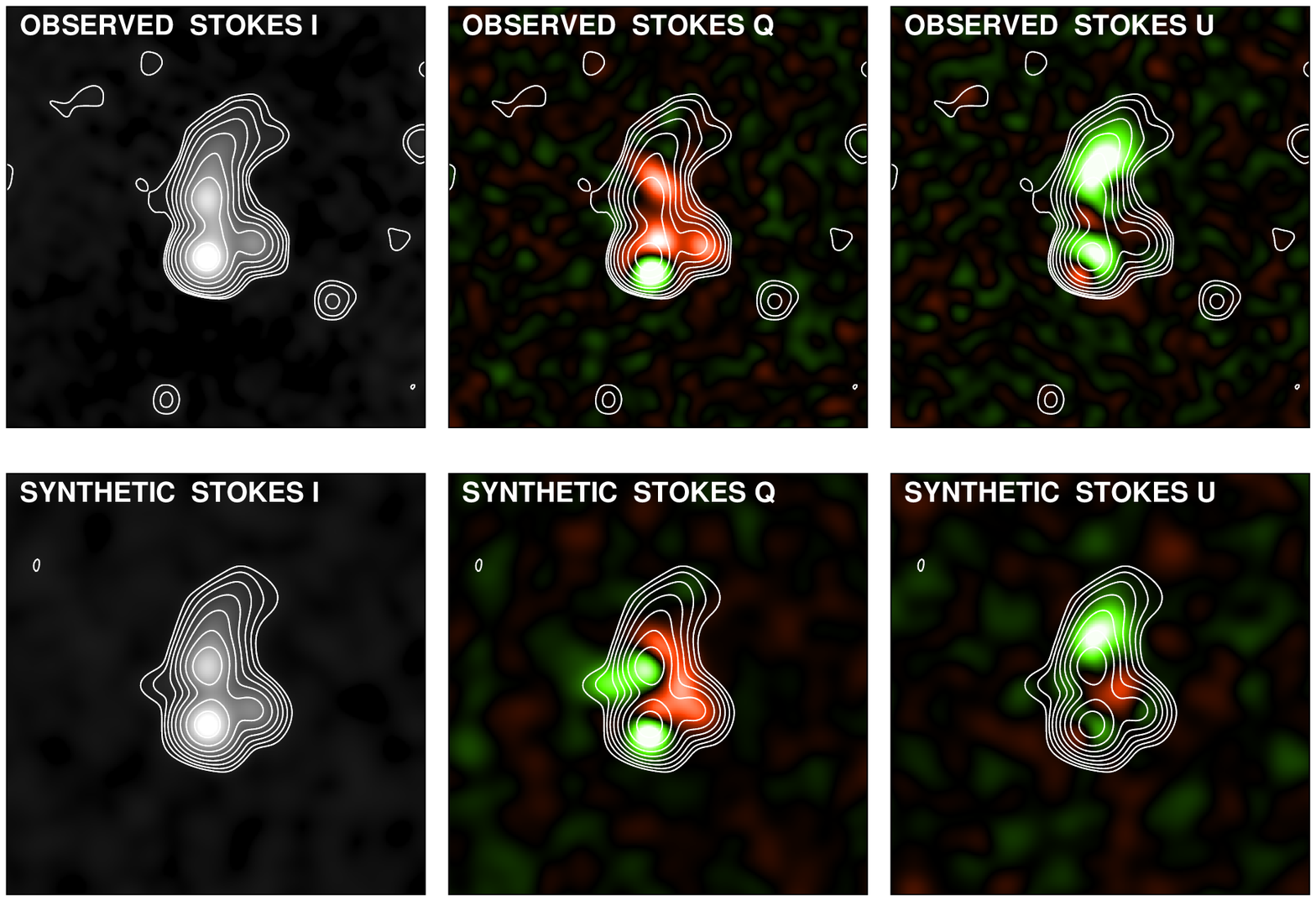}
\caption{Example of observed (top panels) and synthetic (bottom panels) Stokes parameters for the SRT observations at 6600 MHz and 2.9\arcmin resolution. }
 \label{obssynth}
 \end{figure*}

\noindent
Examples of observed and synthetic images used in our analysis are shown in Fig.\,\ref{obssynth}.

Following Vacca et al. (2012), we compared the synthetic images and data using the Bayesian 
inference, whose use was first introduced in the RM analysis by En{\ss}lin \& Vogt
(2003).

 \begin{figure*}[!]
   \centering
  \includegraphics[width=18 cm]{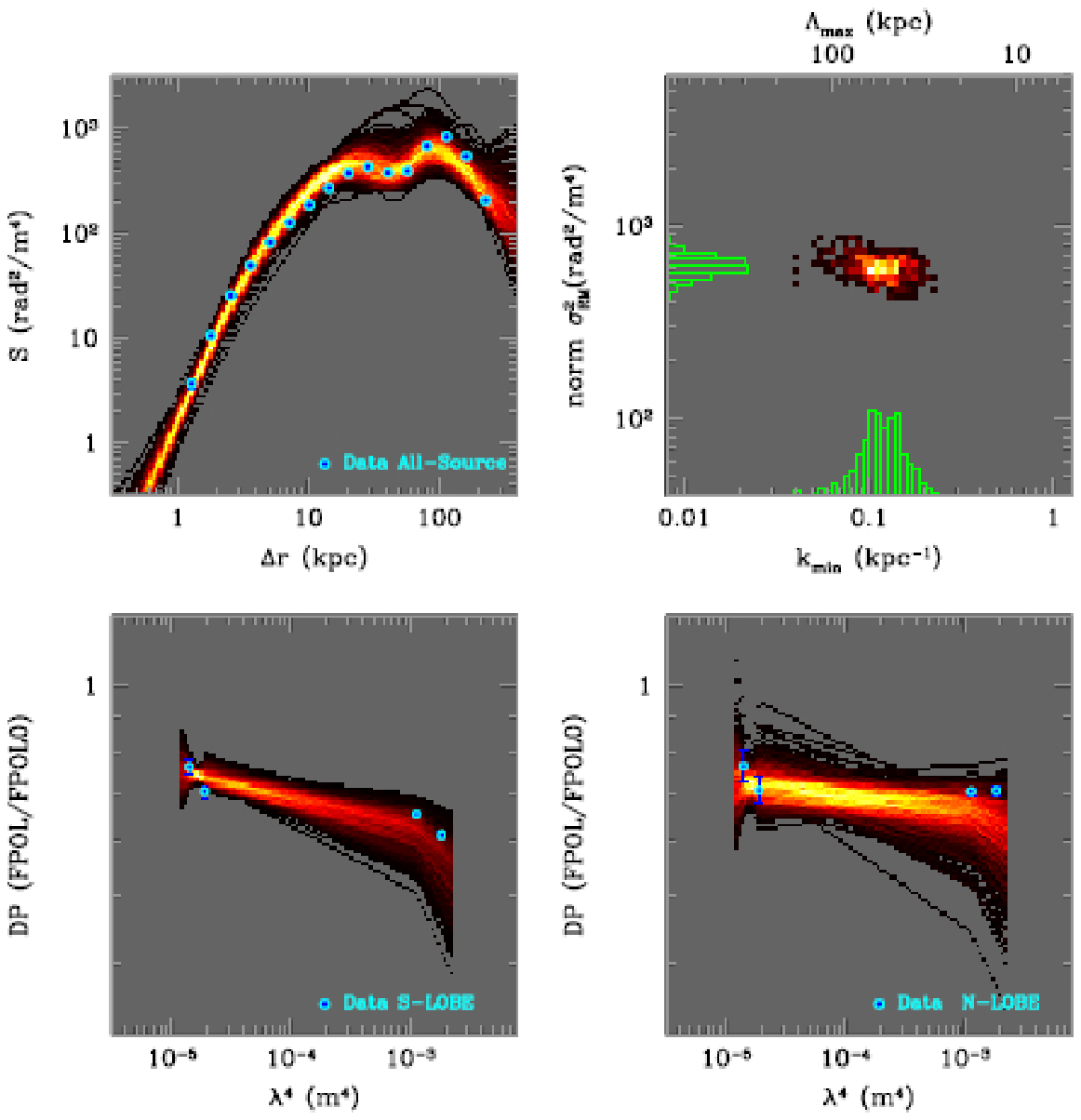}
\caption{Bayesian two-dimensional analysis of the RM structure function $S(\Delta r)$ and
 depolarization ratio FPOL/FPOL$_{\rm 0}(\lambda^4)$. \emph{Top right panel}: image of the posterior distribution of the model parameters along with one-dimensional marginalizations. \emph{Top left panel}:
  The dots represent the observed $S(\Delta r)$ (error bars are comparable to the size of the symbols). 
  The fan-plot represents the population of synthetic RM structure functions from the posterior distribution.
  \emph{Bottom left panel}: Depolarization ratio as a function of $\lambda^4$ for the southern lobe of 3C\,40B.   The dots represent the observed data while the fan-plot represents the population of synthetic depolarization ratio extracted from the posterior distribution of model parameters.  
  \emph{Bottom right panel}: Depolarization ratio as a function of $\lambda^4$ for the northern lobe of 
3C\,40B.
  }
 \label{fit}
 \end{figure*}

\subsection{Bayesian inference}
\label{bayesian}

Because of the random nature of the intra-cluster magnetic field fluctuations, 
the RM and fractional polarization images we observe are just one possible realization of the data.
Thus, rather than try to determine the particular set of power spectrum
parameters, $\vec\theta$, that best reproduces the given realization of the data, it
is perhaps more meaningful to search for that distribution of model
parameters that maximizes the probability of the model given the data. 
The Bayesian inference offers a natural theoretical framework for this approach.

The Bayesian rule relates our prior information on the distribution
$P(\vec\theta)$ of model parameters $\vec\theta$ to their posterior
probability distribution $P(\vec\theta~\vert~D)$ after the data $D$
have been acquired

\begin{equation}
P(  \vec\theta~\vert~D)=\frac{P(D~\vert~ \vec\theta)P(\vec\theta)}{P(D)},
\label{bayes}
\end{equation}

\noindent
where $P (D~\vert~ \vec\theta)$ is the likelihood function, while $P(D)$
is called the evidence. The evidence acts as a normalizing constant
and represents the integral of the likelihood function weighted by the
prior over all the parameters space
\begin{equation}
P(D)=\int P(D~\vert~ \vec\theta)P(\vec\theta)d \vec\theta.
\end{equation} 

The most probable configuration for the model parameters is obtained
from the posterior distribution given by the product of the
likelihood function with the prior probability.
We used a \emph{Markov Chain Monte Carlo} (MCMC) method to extract
samples from the posterior probability distribution. In particular, we
implemented the Metropolis$-$Hastings algorithm, which is capable of
generating a sample of the posterior distribution without the need to
calculate the evidence explicitly, which is often extremely difficult
to compute since it would require exploring the entire parameters space.
The MCMC is started from a random initial position $\vec\theta_{\rm 0}$ and the
algorithm is run for many iterations by selecting new states according
to a transitional kernel, $Q(\vec\theta,\vec\theta^{\rm \prime})$, between
the actual, $ \vec\theta $, and the proposed position,
$\vec\theta^{\rm \prime}$. The proposed position is accepted with
probability

\begin{equation}
h=\min \left[1, \frac{P(D~\vert~  \vec\theta ^{\rm \prime}) P(   \vec\theta^{\rm \prime}) Q(  \vec\theta ^{\rm \prime},  \vec\theta ) } { P(D~\vert~   \vec\theta) P( \vec\theta) Q(  \vec\theta ,  \vec\theta ^{\rm \prime}) } \right].
\end{equation} 

We chose for $Q$ a multivariate Gaussian kernel. The MCMC starts with
a number of ``burn-in'' steps during which, according to common
practice, the standard deviation of the transitional kernel is adjusted so
that the average acceptance rate stays in the range 25\%$-$50\%. After the
burn-in period, the random walk forgets its initial state and the chain
reaches an equilibrium distribution. The burn-in steps are discarded,
and the remaining set of accepted values of $ \vec\theta $ is a
representative sample of the posterior distribution that can be used
to compute the final statistics on the model parameters.

\subsection{2D simulations}
\label{2-dimensional simulations}

The polarization images at a resolution of 19$''$ give a detailed view of the magnetic field
fluctuations toward the brightest parts of the radio sources.
For this reason, we performed a two-dimensional analysis of these data, 
to constrain the maximum scale of the magnetic field fluctuations $\Lambda_{\rm max}$, by fixing
$n$=11/3 and $\Lambda_{\rm min} = 1$ kpc.

Within the power-law scaling regime (En{\ss}lin \& Vogt 2003),
the two-dimensional analysis relies on the proportionality between the magnetic field and
the RM power spectra. On the basis of this proportionality,
the slope $n$ of the two-dimensional RM power spectrum is 
the same of the three-dimensional magnetic field power spectrum:
\begin{equation}
\vert RM_{\rm k}\vert ^{\rm 2}\propto k^{\rm -n}.
\label{rmpower}
\end{equation}

We simulated RM images with a given power spectrum in a
two-dimensional grid. The simulations start in the Fourier space, where
the amplitudes of the RM components are selected according to
Eq.\,\ref{rmpower}, while the phases are completely random. The RM
image in the real space is obtained by a fast Fourier transform (FFT) inversion.
We have taken into account the dependence of the RM on the radial decrease 
of the thermal gas density and of the intra-cluster magnetic field
by applying a tapering to the RM image:
\begin{equation}
RM_{\rm taper}(r)=\left(1+\frac{r^{\rm 2}}{r_{\rm c}^{\rm 2}}\right)^{\rm -3 \beta (1+ \eta)+0.5}
\end{equation}
for which we fixed the 
parameters of the $\beta$-model (Sect.\,\ref{xray}) and we assumed 
in first approximation $\eta=1$.

As a compromise between computational speed and spatial dynamical
range, we adopted a computational grid of 1024$\times$1024\,pixels
with a pixel size of 0.5\,{\rm kpc}. This grid allowed us to explore RM
fluctuations on scales as small as $\Lambda_{\rm min}= 1$\,{\rm kpc},
which is smaller than the linear size of the 19\arcsec~beam of the observations
$FWHM \simeq 6.8$\,{\rm kpc}. At the same time, we were able to investigate
fluctuations as large as $\Lambda_{\rm max}= 512$\,{\rm kpc},
i.e., comparable to the linear size of the radio galaxy 3C\,40B in Abell~194.
 
To the aim of the Bayesian inference outlined in Sect.\,\ref{bayesian},
we adopted a likelihood of the form:
\begin{equation}
P(D~\vert~ \vec\theta)=\prod_{i} \frac{1}{\sigma_{\rm synth}(i)\sqrt{2\pi}} \exp \left[-\frac{1}{2}\frac{\left[D_{\rm obs}(i)-D_{\rm synth}(i)\right]^2}{\sigma_{\rm synth}(i)^2}\right],
\label{boo}
\end{equation}
where the observed data $D_{\rm obs}$ are compared to the synthetic data $D_{\rm synth}$. The synthetic data are obtained by filtering simulations
as the observations (as described in Sect.\,\ref{modeling}). The scatter $\sigma_{\rm synth}$ includes both the simulated noise measurements, $\sigma_{\rm noise}$, and the intrinsic statistical
scatter of the synthetic data, $\sigma D_{\rm synth}$, due to the random nature of the intra-cluster magnetic fields: $\sigma_{\rm synth}^2=\sigma D_{\rm synth}^2+\sigma_{\rm noise}^2$. At each step of the MCMC, for a given combination of the free parameters $\vec\theta$, we performed five simulations with different realizations of the fluctuations phases, and we evaluated the dispersion $\sigma_{\rm synth}$. 

In our two-dimensional analysis, the data $D$ (observed and synthetic) are represented by the azimuthally averaged RM structure function $S(\Delta r)$, and by the depolarization ratio FPOL/FPOL$_{\rm 0}$($\lambda^4$). The posterior distribution for a given set of model parameters is computed by considering the joint 
likelihood of the RM structure function and of the depolarization ratio all together.
We applied the Bayesian method by choosing priors uniform in logarithmic scale for the two free parameters: normalization of the magnetic field power spectrum $norm$ and minimum wave number $k_{\rm min}$.

In the top, right panel of Fig.\,\ref{fit} we show the image of the posterior distribution for the 
two free parameters $norm$ and $k_{\rm min}$.
In addition, one-dimensional marginalizations are shown as histograms along
each axis of the image. The one-dimensional marginalizations represent the projected density of samples in the MCMC, which is
proportional to the posterior probability of that particular couple of
model parameters. 

To provide a visual comparison between synthetic and observed data we present the fan-plots of the 
structure function and of the depolarization ratio.

In the top left panel of Fig.\,\ref{fit} the dots represent the observed RM structure function $S(\Delta r)$ 
calculated using the RM image of Fig.\,\ref{Fig_MidRes}. 
The observed RM structure function $S(\Delta r)$ is shown together with the population of synthetic
structure functions extracted from the posterior distribution. Brighter
pixels in the fan-plot occur where many different synthetic
structure functions overlap each other. 
Thus, the fan-plot gives a visual indication of the marginalized posterior probability of the synthetic data at 
a given separation.
Overall, the fan-plot shows that the Kolmogorov power spectrum is able to 
reproduce the shape of the observed RM structure function, including the large scale oscillation observed before 
the turnover at $\Delta r>100$ kpc. 
The turnover is likely caused by under-sampling of the large separations imposed by the window function due to the silhouette of the radio sources combined with the effect of the tapering imposed by the dimming of the gas density and magnetic field strength at increasing distances from the cluster center. On small separations,
 $\Delta r< 6.8$ kpc, the slope of the structure function is affected by the observing beam. It is worth mentioning 
that all these effects are fully taken into account when simulating the synthetic I, Q, and U images.

In the bottom panels of Fig.\,\ref{fit} the dots represent
FPOL/FPOL$_{\rm 0}$ as a function of $\lambda^4$ calculated in the southern (left panel) and in the northern
(right panel) lobe of 3C\,40B, respectively.
The southern lobe, is slightly more depolarized than the norther lobe. 
The statistics of the fractional polarization have been calculated 
considering only the pixels in which FPOL$>$3$\sigma_{\rm FPOL}$.
The observed  depolarization ratio as a function of $\lambda^4$ is shown, together with the
fan-plots of the synthetic depolarization. 
In general, the Kolmogorov power spectrum is able to reproduce the observed
depolarization, as shown by the fan-plots.

To summarize, the combined two-dimensional analysis of RM structure function and
fractional polarization allowed us to constrain the maximum scales on the magnetic 
field fluctuations to $\Lambda_{\rm max}$=(64$\pm$24)\,{\rm kpc}, where the given error represents
the dispersion of the one-dimensional marginalizations.

The magnetic field auto-correlation length is calculated according to:

\begin{equation}
\Lambda_{\rm B}=\frac{3\pi}{2}\frac{\int_{0}^{\infty}|B_{k}|^2k dk}{\int_{0}^{\infty}|B_{k}|^2 k^2 dk},
\end{equation}
where the wavenumber $k=2\pi/\Lambda$ (En{\ss}lin \& Vogt 2003). We obtained $\Lambda_{\rm B}$=(20$\pm$8)\,{\rm kpc}.

In the next step we fix this value and we constrain the strength of the magnetic field and its scaling
with the gas density with the aid of three-dimensional simulations.

\begin{figure*}[ht]
  \centering
  \includegraphics[width=15 cm]{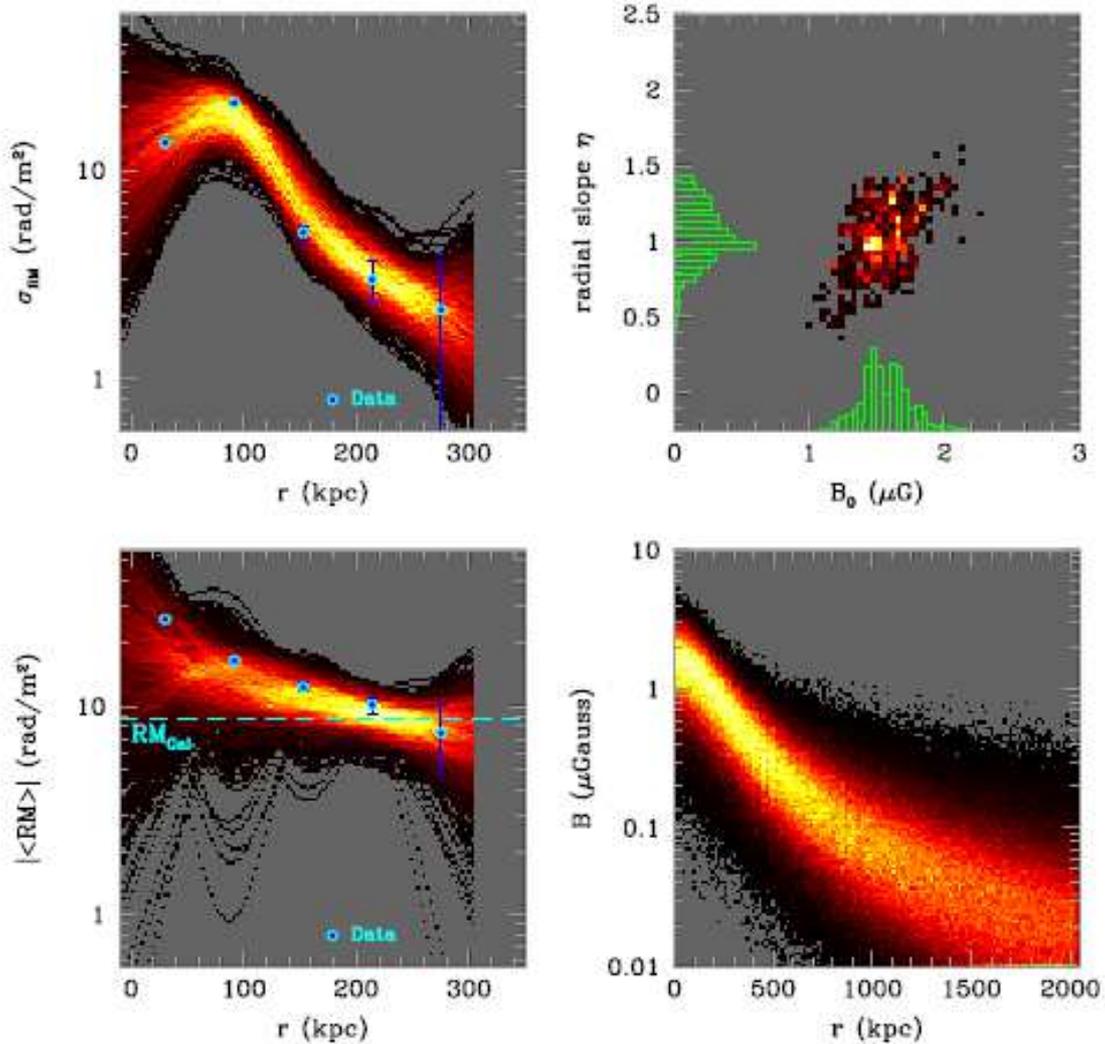}
  \caption{Bayesian three-dimensional analysis of the $\sigma_{RM}$ and $\langle RM \rangle$ as a function of the distance from the cluster center. 
 \emph{Top right panel}: image of the posterior distribution of the model parameters along with one-dimensional marginalizations.
\emph{Top left panel}: The dots represent the observed  $\sigma_{RM}$ profile as a function of the 
   distance from the cluster center while the fan-plot represents the population of synthetic  $\sigma_{RM}$ profile extracted from the posterior 
distribution of model parameters.
\emph{Bottom left panel}: The dots represent the observed $\langle RM \rangle$ profile as a function of the 
   distance from the cluster center while the fan-plot represents the population of synthetic $\langle RM \rangle$ profile extracted from the posterior 
distribution of model parameters. The Galactic foreground of 8.7 rad/m$^2$ (Oppermann et al. 2015), is indicated with a dashed line.
 \emph{Bottom right panel}: Trend of the mean magnetic field with the distance from 
the cluster center.
}
  \label{Btrend}
\end{figure*}

\subsection{3D simulations}
\label{The magnetic field model}

The data at a resolution of 2.9$'$ are best suited to derive RM information 
up to a large distance from the cluster center. Therefore, 
we used these data to constrain the strength of the magnetic field and
its scaling with the gas density using three-dimensional simulations.

We constructed three-dimensional models of the intra-cluster magnetic
field by following the numerical approach described in Murgia et
al. (2004). The simulations begin in Fourier space by extracting the
amplitude of the magnetic field potential vector, $\tilde A(k)$, from
a Rayleigh distribution whose standard deviation varies with the wave
number according to $|A_{\rm k}|^{\rm 2}\propto k^{\rm -n-2}$.  The phase of the
potential vector fluctuations is taken to be completely random. The
magnetic field is formed in Fourier space via the cross product
$\tilde B(k)=i k \times \tilde A(k)$. This ensures that the magnetic
field is effectively divergence free.  We then perform a three-dimensional 
FFT inversion to produce the magnetic field in the
real space domain.  The field is then Gaussian and isotropic, in the
sense that there is no privileged direction in space for the magnetic
field fluctuations. The power-spectrum normalization is set such that
the average magnetic field strength scales as a function of the
thermal gas density according to Eq.\,\ref{eta}. The gas density
profile is described by the $\beta$-model in Eq.\,\ref{beta}, 
modified, as done in A2199 by Vacca et al. 2012, to take into account 
of the presence of the X-ray cavity detected in correspondence of the 
south lobe of 3C\,40B.

We simulated the random magnetic field by using a cubic grid of
1024$^{\rm 3}$\, cells with a cell size of 0.5\,{\rm kpc}/cell. 
The simulated magnetic
field is periodic at the boundaries and thus the computational grid is replicated to
cover a larger cluster volume.

\begin{table*}[h]
\caption{Cluster magnetic field values taken from the literature.}
\begin{center}
\begin{tabular} {lccll} 
\hline
Cluster    & $\langle  B_{0}\rangle$    &     $n_0$        &  T           & References      \\
           & ($\mu$G)                  &    (cm$^{-3}$)    &  (keV)       &                 \\
\hline
Abell~194      &        1.5                &  0.69          & 2.4         & This work       \\
Abell~119      &        5.96               &  1.40          & 5.6         & Murgia et al. (2004)         \\
Abell~2199     &        11.7               & 101.0          & 4.1         & Vacca et al. (2012)         \\
Abell~2255     &        2.5                &  2.05          & 6.87        & Govoni et al. (2006)         \\
Abell~2382     &       3.6                 & 5.0            & 2.9            & Guidetti et al. (2008)       \\
3C31           &        6.7                &  1.9           & 1.5      & Laing et al. (2008)        \\
3C449          &        3.5                &     3.7        & 0.98     & Guidetti et al. (2010)     \\
Coma           &        4.7                & 3.44           & 8.38     & Bonafede et al. (2010)         \\
Hydra          &        45.2               & 62.26          & 4.3            & Laing et al. (2008)         \\
\hline
\multicolumn{5}{l}{\scriptsize Col. 1: Cluster; Col. 2: Central magnetic field; Col. 3: Central electron density;}\\ 
\multicolumn{5}{l}{\scriptsize References for $n_0$, converted to our cosmology, are taken from:}\\ 
\multicolumn{5}{l}{\scriptsize This work (Abell~194); Cirimele et al. 1997 (A119); Johnstone et al. 2002 (Abell~2199);}\\  
\multicolumn{5}{l}{\scriptsize Feretti et al. 1997 (Abell~2255); Guidetti et al. 2008 (Abell~2382);}\\ 
\multicolumn{5}{l}{\scriptsize Komossa \& B{\" o}hringer 1999 (3C31); Croston et al. 2008 (3C449);}\\
\multicolumn{5}{l}{\scriptsize Briel et al. 1992 (Coma); Wise et al. 2007 (Hydra).}\\
\multicolumn{5}{l}{\scriptsize Col. 4: Temperature; References for T are taken from:}\\
\multicolumn{5}{l}{\scriptsize This work (Abell~194); Reiprich \& B{\" o}hringer 2002 (Abell~119, Abell~2255, Abell~2199, Coma, HydraA);}\\
\multicolumn{5}{l}{\scriptsize Ebeling et al. 1996 (Abell~2382); Croston et al. 2008 (3C31, 3C449).}\\
\multicolumn{5}{l}{\scriptsize Col. 5: References for $\langle B_{0}\rangle$.}\\ 
\multicolumn{5}{l}{\scriptsize For the comparison with the results by Laing et al. (2008), we converted B$_{\rm rms}$ to $\langle  B_{\rm 0}\rangle$.}\\ 
\multicolumn{5}{l}{\scriptsize In the case of 3C\,31, we report the magnetic field calculated without including X-ray cavities.}\\  
\end{tabular}
\label{letteratura}
\end{center}
\end{table*}

\begin{figure*}[h]
\centering
\includegraphics[width=17 cm]{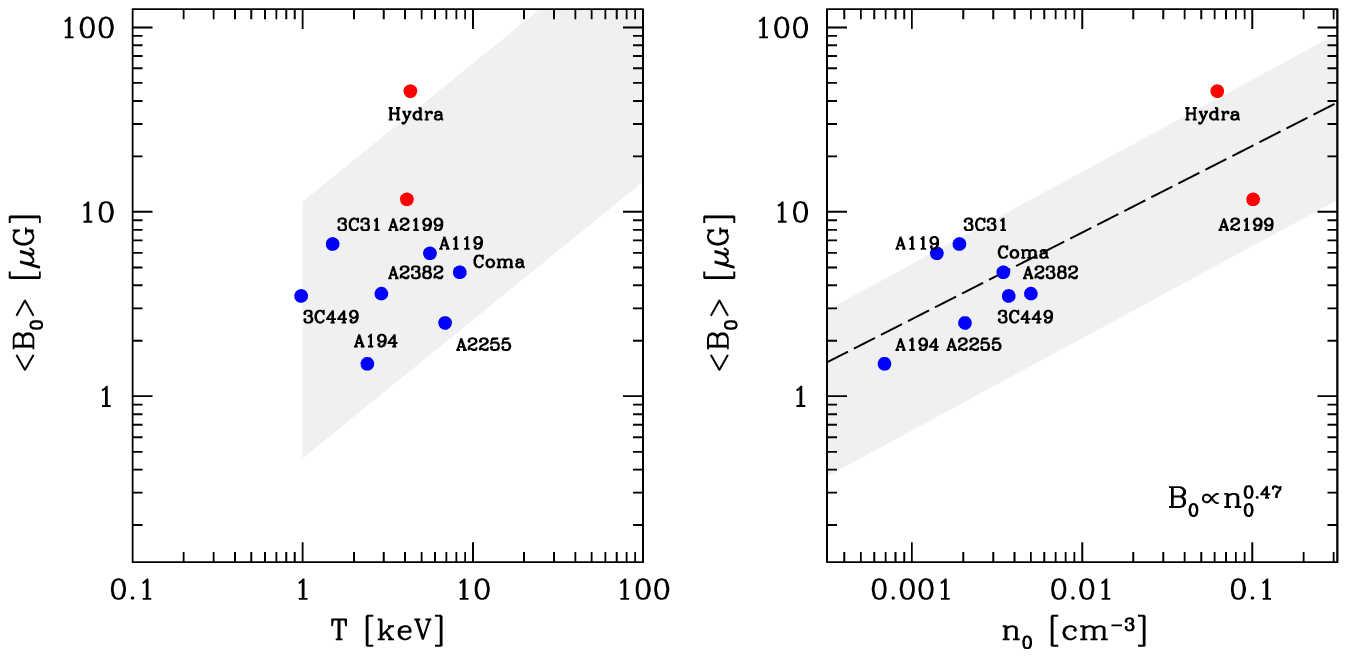}
\caption[]{
 \emph{Right panel}: Plot of the central magnetic field strength  
$\langle  B_{\rm 0}\rangle$ versus the mean cluster temperature. Red dots
indicate cooling core clusters. 
\emph{Left panel}: Plot of  $\langle  B_{\rm 0}\rangle$ versus the central electron density $n_0$.
Red dots indicate cooling core clusters. 
Dashed line indicates the scaling obtained by a linear fit of the log-log relationship: 
$\langle  B_{\rm 0}\rangle$ $\propto$ $n_0^{0.47}$.
The shaded regions represent the expected range of theoretical field strengths expected on the basis of the
modeling by Kunz et al. (2011). 
}
\label{bmean}
\end{figure*}

The simulated RM images were obtained by numerically integrating 
the magnetic field and the gas density product along the line-of-sight, 
accordingly to
Eq.\,\ref{rm}. In the case of Abell~194, the integration was performed from the
cluster centre up to 1 Mpc along the line-of-sight.
In a similar way to the two-dimensional simulations, the simulated RM images were filtered 
like the observations. Finally, we used the Bayesian approach to find the posterior distribution for
$\langle B_{\rm 0}\rangle$ and $\eta$. We derived the observed profiles of 
 $\sigma_{RM}$ and $\langle RM \rangle$ as a function of the distance from the cluster center
 by calculating the statistics in five concentric annuli of 2.9\arcmin\, in size. The annuli are
centered at the position of the X-ray centroid (see Sect.\,\ref{xray}).

The slope $n=11/3$ and the scales $\Lambda_{\rm min}=1$ kpc 
and $\Lambda_{\rm max}=64$ kpc were kept fixed at the values found in the two-dimensional analysis.

We applied the Bayesian method by choosing an uniform prior for the distribution of $\eta$, and a 
prior uniform in logarithmic scale
for the distribution of $\langle B_{\rm 0}\rangle$. At each step of the MCMC we simulated the 
synthetic I, Q, and U images of 3C\,40A and 3C\,40B and we derived the synthetic RM image following
the same procedure as for the two-dimensional analysis. Using Eq.\,15, we evaluated the likelihood of the $\sigma_{RM}$ 
and $\langle RM \rangle$ profiles by mean of ten different random configurations for the magnetic field phases to
calculate the model scatter $\sigma D_{\rm synth}$.

The result of the Bayesian analysis is shown in Fig.\,\ref{Btrend}. 
In the top right panel of Fig.\,\ref{Btrend}, we present the image of the posterior distribution of
model parameters along with the one-dimensional marginalizations represented as histograms. 

In the top left panel of Fig.\,\ref{Btrend} we show the observed $\sigma_{RM}$ profile 
along with the fan-plot of the synthetic profiles from the posterior. The dispersion of the
RM increases toward the cluster centre as expected due to the higher density and magnetic 
field strength. The drop at the cluster center is likely caused by the 
under-sampling of the RM statistics as a result of the smaller size of the central annulus with respect to the external annuli.
In the bottom left panel of Fig.\,\ref{Btrend} we show the observed $\langle RM \rangle$ profile 
along with the fan-plot of the synthetic RM profiles extracted from the posterior distribution.
In this plot it is evident that the cluster shows an enhancement of the $\langle RM \rangle$
signal with respect to the Galactic foreground of 8.7 rad/m$^2$ (Oppermann et al. 2015).

Finally, in the bottom right panel of Fig.\,\ref{Btrend} we show the trend of the mean magnetic field with the distance from 
the cluster center.

The degeneracy between the central magnetic field strength and radial index is evident in the posterior
distribution of model parameters: the higher $\langle B_{\rm0}\rangle$ the
higher $\eta$. However, we were
able to constrain the central magnetic field strength to $\langle B_{\rm0}\rangle$=(1.5$\pm$0.2)\,$\mu$G, 
and the radial index to $\eta$=(1.1$\pm$0.2). 
The average magnetic field over a volume of 1 Mpc$^{3}$ is as 
low as $\langle B \rangle \simeq$ 0.3 $\mu$G. 
This result is obtained by fixing the slope of the magnetic field power spectrum to the
Kolmogorov index. However, we note that the central magnetic field strength depends on the square root of the 
field correlation length $\Lambda_{B}$ (see Eq.\,16). The field correlation length in turn depends on a combination
of the power spectrum slope and $\Lambda_{max}$ and these two parameters are degenerate: a flatter slope for instance leads
to a larger $\Lambda_{max}$ so that $\Lambda_{B}$ is preserved. Therefore, a different choice of the power spectrum slope $n$
should have a second-order effect on the estimated central magnetic field strength. Nevertheless, as a further check,
we repeated the 2D Bayesian analysis (not shown) by freeing all the power spectrum parameters: 
normalization, $\Lambda_{min}$, $\Lambda_{max}$, and $n$. The larger dimensionality implies larger uncertainties, due to 
the computational burden needed to fully explore the parameters space. The maximum posterior probability is found 
for a slope $n\simeq 3.6\pm 1.6$, which is in agreement, to within the large uncertainty, with the Kolmogorov slope $n=11/3$.

To our knowledge, the central magnetic field we determined in Abell~194 is the 
weakest ever found using RM data in galaxy clusters. We note, however, 
that Abell~194 is a poor galaxy cluster with no evidence of cool core.

It is interesting to compare the intra-cluster magnetic field of Abell~194 to that of other galaxy clusters for which 
a similar estimate is present in the literature. These are listed in Table\,\ref{letteratura}. 

In the left panel of Fig.\,\ref{bmean} we show a plot of the central magnetic field strength  
$\langle  B_{\rm 0}\rangle$ versus the mean cluster temperature, while 
in the right panel of Fig.\,\ref{bmean} we show a plot of       
$\langle  B_{\rm 0}\rangle$ versus the central electron density $n_0$.

Although the cluster sample is still rather small, and thus all the following considerations should be taken with care, 
we note that there is a hint of a positive trend between   
$\langle  B_{\rm 0}\rangle$ and $n_0$ measured among different clusters.
On the other hand, no correlation seems to be present between the central magnetic field and the mean cluster temperature,
in agreement with what found by Govoni et al. (2010) 
in a statistical analysis of a sample of RM in rich galaxy clusters. 

There are three merging clusters in the sample: Coma, A2255, and A119. They all have 
quite similar central magnetic field strengths despite Coma and A2255 host a giant radio halo while A119 is radio quite.
We confirm that cool core clusters like Hydra and A2199 tend to have
higher central magnetic fields. In general, fainter central 
magnetic fields seems to be present in less dense galaxy clusters. 
Actually, this result is corroborated by the low central magnetic field found in this work for the 
poor galaxy cluster Abell~194. The scaling obtained by a linear fit of the log-log relationship 
is: $\langle  B_{\rm 0}\rangle$ $\propto$ $n_0^{0.47}$.
This result is in line with the theoretical prediction by Kunz et al. (2011), who assume that parallel viscous heating due to turbulent
dissipation balances radiative cooling at all radii inside the cluster core. 
Kunz et al. (2011) found in the bremsstrahlung regime (that is $T\gtrsim$ 1\,keV)
\begin{equation}
B\simeq 11\xi^{\rm -1/2}\left(\frac{n_{\rm e}}{0.1{\rm cm}^{\rm -3}}\right)^{\rm 1/2}\left(\frac{T}{2{\rm keV}}\right)^{\rm 3/4} \mu {\rm G},
\label{kunz}
\end{equation}
where $T$ is the temperature, while $\xi$ is expected to range between
0.5 and 1 in a turbulent plasma.
Following this formula, the expected magnetic field at the center of Abell~194
is 1.05$\xi^{-1/2}\mu$G.
Although the central magnetic field found in our analysis is in remarkable agreement with the value obtained from Eq.\,\ref{kunz},
we note that we found a magnetic field which scales with density as $\eta=1.1\pm0.2$, while according to
Eq.\,\ref{kunz}, we should expect $\eta=0.5$. However, we did not included in our analysis a possible scaling of $B$ with the cluster
temperature. If the intra-cluster medium is not isothermal, it is possible to bring the magnetic field radial profile
found in our analysis in agreement with that in Eq.\,\ref{kunz} supposing in Abell~194
a temperature profile as steep as $T\propto r^{-1.4}$ at large radii (r$>$ 0.2 $R_{180}$). The temperature profile for Abell~194
is not known at large radii, but the required temperature profile is much steep
than the typical temperature decreases found in galaxy clusters which is
$T\propto r^{\mu}$ with $\mu$ between
0.2 and 0.5 (e.g. Leccardi \& Molendi 2008).

In Fig.\,\ref{bmean}, we plot the expected range of magnetic field strengths obtained by using Eq.\,\ref{kunz}.
In the right panel, the upper bound corresponds to $\xi$=0.5 and T=10 keV, while the lower bound corresponds
to  $\xi$=1 and T=1 keV.
In the left panel, the upper bound corresponds to $\xi$=0.5 and $n_0=5\times10^{-4}$ cm$^{-3}$, while the lower bound corresponds
to  $\xi$=1 and $n_0=0.15$ cm$^{-3}$. Indeed for some reason it seems that the model by Kunz et al. (2011) seems
to reproduce the scaling between central magnetic field and gas density among different galaxy clusters
even if in the case of Abell~194 the observed magnetic field scaling with radius is significantly steeper than what they predict.

\section{Summary and Conclusions}

In the context of the SRT early science program SMOG
(SRT Multifrequency Observations of Galaxy clusters),
we have presented a spectral-polarimetric study of 
the poor galaxy cluster Abell~194.
We have analyzed total intensity and polarization images of the extended radio galaxies 3C\,40A and 3C\,40B located close to the cluster center.
By complementing the SRT observations with radio data at higher resolution and at different
frequencies, but also with data in optical and X-ray bands,
we aimed at inferring the dynamical state of the cluster, the radiative age of the radio galaxies, and the 
intra-cluster magnetic field power spectrum. The SRT data were used in combination with archival 
Very Large Array observations at lower frequencies to derive both spectral aging and RM images of 3C\,40A and 3C\,40B.
The break frequency image and the cluster electron density profile inferred from \rosat\ and \chandra\ observations 
are used in combination with the RM data to constrain the intra-cluster magnetic field power spectrum. 
Following Murgia et al. (2004), 
we simulated Gaussian random two-dimensional and three-dimensional magnetic
field models with different power-law power spectra, and we compared
the observed data and the synthetic images with a 
Bayesian approach (Vogt \& En{\ss}lin 2005, Vacca et al. 2012), in order to constrain the strength 
and structure of the magnetic field associated with the intra-cluster medium.
Our results can be summarized as follows:\\

$-$  To investigate the dynamical state of Abell~194 and to obtain 
new additional insights in the cluster structure, we analyzed 
the redshifts of 1893 galaxies from the literature, resulting in a sample of 143 fiducial cluster members.
The picture resulting from new and previous results agrees in that Abell~194 does not
show any trace of a major and recent cluster merger, but rather agrees with a scenario of accretion of
small groups, mainly along the NE-SW axis.\\ 

$-$ The measured break frequency in the radio spectra decreases systematically along the lobes of 
3C\,40B, down to a minimum of 850$\pm$120 MHz. If we assume that the magnetic field
intrinsic to the source 3C\,40B is in equipartition,
the lifetimes of radiating electrons result 157$\pm$11 Myrs, in agreement with the 
spectral age estimate calculated by Sakelliou et al. (2008). 
Furthermore, the radiative age of 3C40\,B is consistent 
with that found in the literature for other sources of similar size and radio power (Parma et al. 1999).\\

$-$ Linearly polarized emission is clearly detected for both 
sources and the resulting polarization angle images are used to produce detailed
RM images at different angular resolutions (2.9\arcmin, 60\arcsec, and 19\arcsec).
The RM images of Abell~194 show patchy patterns without any obvious
preferred direction, in agreement with the standard picture that the Faraday effect is due to the
turbulent intra-cluster medium.\\
 
\begin{table}
\caption{Results of the intra-cluster magnetic field power spectrum in Abell 194.}    
\centering          
\begin{tabular}{l c} 
\hline
Parameter                & Value         \\ 
\hline                        
  $n$ & 11/3~~(fixed)  \\
 $\Lambda_{\rm min}$ & 1 kpc~~(fixed)  \\
 $\Lambda_{\rm max}$ & 64$\pm$24 kpc \\
 $\Lambda_{\rm B}$ & 20$\pm$8 kpc \\
 $\langle  B_{\rm 0}\rangle$ &1.5$\pm$0.2 $\mu$G\\
 $\eta$ & 1.1$\pm$0.2  \\
\hline  
\multicolumn{2}{l}{\scriptsize Col. 1: Parameters of the magnetic field model;}\\ 
\multicolumn{2}{l}{\scriptsize Col. 2: Results of the magnetic field parameters.}\\ 
\end{tabular}
\label{results} 
\end{table}

$-$ The results of the magnetic field power spectrum in Abell~194 are reported in Table\,\ref{results}.
By assuming a Kolmogorov magnetic field power law power spectrum (slope $n$=11/3) with a minimum scale of 
fluctuations $\Lambda_{min}=1$ kpc, we find that the RM data 
in Abell~194 are well described by a magnetic field with a maximum scale of fluctuations of 
$\Lambda_{max}=(64\pm24)$ kpc. 
The corresponding magnetic field auto-correlation length is $\Lambda_{\rm B}$=(20$\pm$8) kpc.
We find a magnetic field strength of $\langle B_{\rm 0}\rangle$=(1.5$\pm$0.2) $\mu$G at the cluster center. 
To our knowledge, the central magnetic field we determined in Abell~194 is the weakest ever 
found using RM data in galaxy clusters. We note, however, that Abell~194 is a poor galaxy 
cluster with no evidence of cool core.
Our results also indicate a magnetic field strength which scales as
the thermal gas density. In particular,
the field decreases with the radius following the gas density to the power 
of $\eta$=(1.1$\pm$0.2).\\

$-$ We compared the intra-cluster magnetic field determined for Abell~194 
to that of a small sample of galaxy clusters for which a similar estimate was available in the literature.
No correlation seems to be present between the mean central magnetic field and the cluster temperature. 
On the other hand, there is a hint of a trend between the central electron densities and magnetic field 
strengths among different clusters: $\langle  B_{\rm 0}\rangle$ $\propto$ $n_0^{0.47}$.

\begin{acknowledgements}
We thank the anonymous referee for the useful comments who helped to improve the paper.\\
The Sardinia Radio Telescope is funded by the Department of University and Research (MIUR),
Italian Space Agency (ASI), and the Autonomous Region of Sardinia (RAS) 
and is operated as National Facility by the National Institute for Astrophysics (INAF).\\
The development of the SARDARA back-end has been funded by the Autonomous Region of Sardinia (RAS)
using resources from the Regional Law 7/2007 "Promotion of the scientific research and
technological innovation in Sardinia" in the context of the research project CRP-49231 (year
2011, PI Possenti): "High resolution sampling of the Universe in the radio band: an unprecedented
instrument to understand the fundamental laws of the nature".\\ 
This research was partially supported by PRIN-INAF 2014.\\
The National Radio Astronomy Observatory (NRAO)
is a facility of the National Science Foundation, operated under
cooperative agreement by Associated Universities, Inc.
This research made use of the NASA/IPAC Extragalactic Database (NED) which is operated by
the Jet Propulsion Laboratory, California Institute of Technology, 
under contract with the National Aeronautics and Space Administration.\\ 
F. Loi gratefully acknowledges Sardinia Regional Government for the
financial support of her PhD scholarship 
(P.O.R. Sardegna F.S.E. Operational Programme of the Autonomous Region of Sardinia, European
Social Fund 2007-2013 - Axis IV Human Resources, Objective l.3, Line of Activity l.3.1.).\\
Basic research in radio astronomy at the Naval Research Laboratory is funded
by 6.1 Base funding.\\
This research was supported by the DFG Forschengruppe 1254
“Magnetisation of Interstellar and Intergalactic Media: The Prospects of
Low-Frequency Radio Observations”.\\
F. Vazza acknowledges funding from the European Union's Horizon 2020 research and innovation programme 
under the Marie-Sklodowska-Curie gran agreement No 664931.
\end{acknowledgements}

\end{document}